\documentclass[aps,prb,longbibliography,twocolumn,amsmath,amssymb]{revtex4-1}

\usepackage{graphics}
\usepackage{graphicx}
\usepackage{bm}
\usepackage{natbib} 
\usepackage{amsmath}
\usepackage{mathtools}
\usepackage{float}
\usepackage{verbatim}

\showboxbreadth\maxdimen\showboxdepth\maxdimen

\begin{document}

\raggedbottom

\title{Introduction to molecular dynamics simulations}

\author{K. Vollmayr-Lee}
\affiliation{Department of Physics and Astronomy, Bucknell University, 
Lewisburg, Pennsylvania 17837}

\begin{abstract}
We provide an introduction to molecular dynamics simulations 
in the context of the 
Kob-Andersen model of a glass.
We introduce a complete set of tools for doing and analyzing the results 
of simulations at fixed NVE and NVT. The modular 
format of the paper  allows
readers to select sections that meet their needs.
We start with an introduction to molecular dynamics  
independent of the programming language, followed by introductions 
to an implementation using Python and then the freely available open source 
software package LAMMPS. We also describe analysis tools for the quick 
testing of the program during its development and compute the radial distribution
function and the mean square displacement
using both Python and LAMMPS.
\end{abstract}

\maketitle

\section{Introduction}
\label{sec:intro}
Computer simulations are a powerful 
approach for addressing questions which are not accessible by theory 
and experiments. Simulations give 
us access to analytically unsolvable systems,
and contrary to laboratory experiments, there are no unknown ``impurities'' 
and we can work with a well defined model.
In this paper we focus on the simulation 
of many particle systems using molecular dynamics,
which models a system of classical particles whose dynamics is described 
by Newton's equations and its generalizations.

Our goal is to provide 
the background for those 
who wish to use and analyze molecular 
dynamics simulations. This paper may also be used 
in a computer simulation course or for  student projects 
as part of a course.

Many research groups no longer write
their own molecular dynamics programs, but use instead 
highly optimized and  complex software 
packages such as LAMMPS.\cite{LAMMPSwebpage} To understand the core of 
these software packages and how to use them wisely, it 
is educational for students to write and use
their own program before continuing with a software package. One intention of this paper is to guide students through an
example of a molecular dynamics simulation and then 
 implement the same task with LAMMPS. 
A few examples are given to illustrate the wide variety
of possibilities for analyzing molecular dynamics
simulations and hopefully to lure students into  
investigating the beauty of many particle systems.

Although we discuss the Python programming language, the 
necessary tools are independent of the programming language and are introduced in Sec.~\ref{sec:MDSimulation}. Those who prefer to start programming 
with minimal theoretical background may 
start with Sec.~\ref{sec:MDPython} and follow the guidance provided 
on which subsection of Sec.~\ref{sec:MDSimulation}
is important for understanding 
the corresponding  subsection in Sec.~\ref{sec:MDPython}. 

Throughout the paper we refer to problems that are listed in Sec.~\ref{sec:problems}.
Answers to  Problems (\ref{prob:Fix})--(\ref{prob:Hderivation}) 
are given in the text immediately following 
their reference.  
These suggested problems are intended to encourage active engagement with the paper by encouraging 
readers to work out sections of the paper by themselves.  

\section{Molecular Dynamics Simulation}
\label{sec:MDSimulation}

\subsection{Introduction}
\label{sec:MDSimIntro}

\begin{figure}[t]
\begin{center}
\includegraphics[width=\columnwidth]{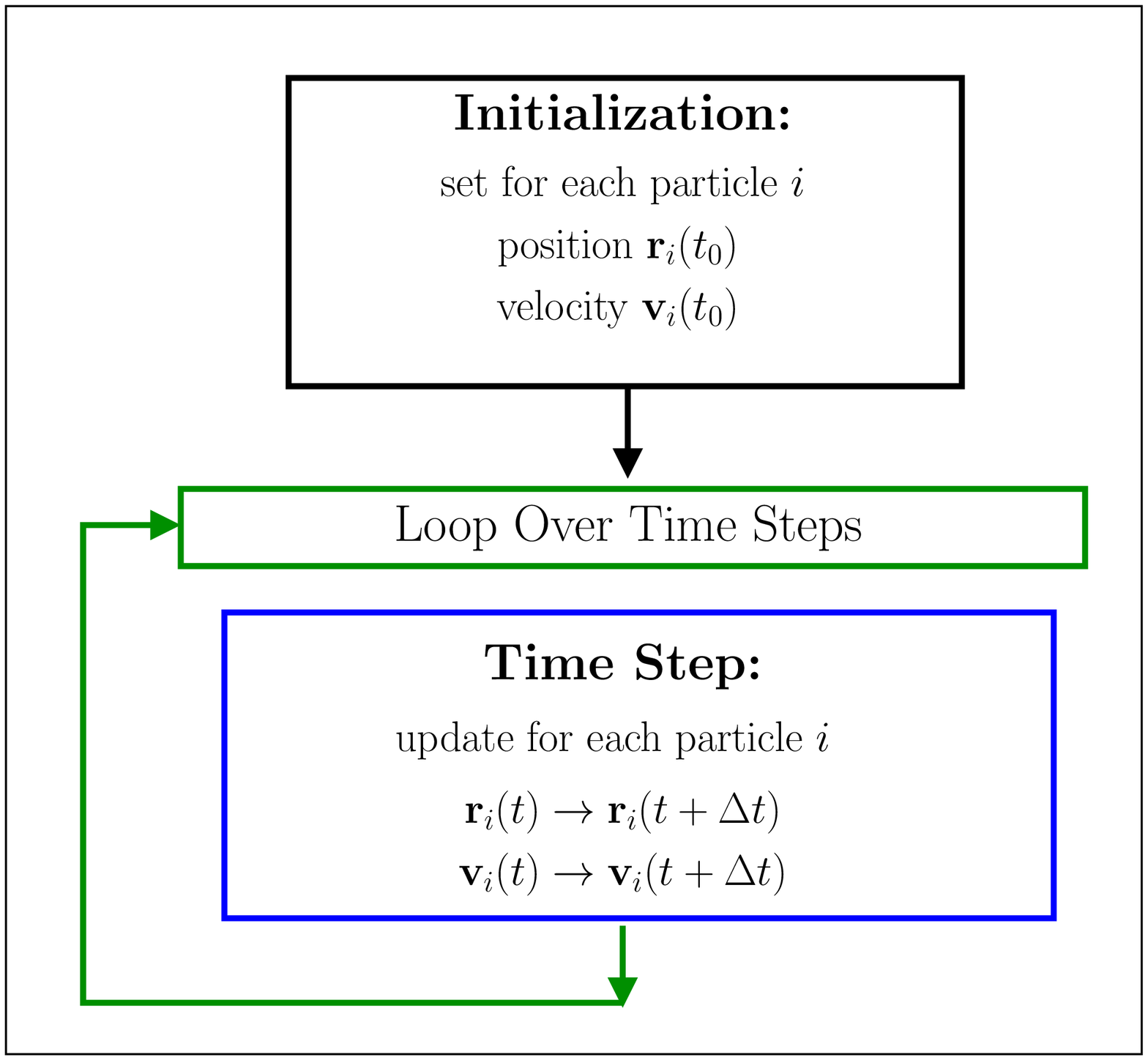}
\caption{Flow chart of a molecular dynamics simulation 
program.} 
\label{fig:MDintro}
\end{center}
\end{figure}

Molecular dynamics simulates a classical system 
of $N$ particles.
The core of most simulations is 
to start with the initial positions and velocities
of all particles and to then repeatedly apply a ``recipe'' to update 
each particle's position and velocity from time $t$ to time
$t+\Delta t$ (see Fig.~\ref{fig:MDintro}). The dynamics 
is governed by Newton's second law
\begin{equation}
 \label{eq:Newton2ndlaw}
 \mathbf{F}_i = m_i \mathbf{a}_i,
\end{equation}
where $\mathbf{a}_i$ is the acceleration of particle $i$.

In Sec.~\ref{sec:system} we define the net 
force $\mathbf{F}_i$ which is used in this paper
and discuss in Sec.~\ref{sec:MDstep} 
the  update rules for the positions $\mathbf{r}_i$ and velocities $\mathbf{v}_i$.

\subsection{Model}
\label{sec:system}

The model is specified by the net force $\mathbf{F}_i$ 
on each particle of mass $m_i$. The force $\mathbf{F}_i$ can be
due to all other particles and/or additional interactions 
such as effective drag forces or interactions with 
a wall or an external field. In the following we will consider
only conservative forces which are due to all the other particles. We also assume 
pair-wise interactions given by a potential 
\begin{equation}
 \label{eq:Vij}
 V = \sum \limits_{i=1}^{N-1} \sum \limits_{j=i+1}^N V_{ij}.
\end{equation}
Specifically, we use 
the Lennard-Jones potential
\begin{equation}
\label{eq:VLJ}
V_{ij} = 4 \epsilon \left [ 
 \left ( \frac{\sigma}{r_{ij}} \right )^{12}
- \left ( \frac{\sigma}{r_{ij}} \right )^{6} 
 \right ],
\end{equation}
where $r_{ij} = | \mathbf{r}_i - \mathbf{r}_j|$ is the distance
between particle $i$ at position $\mathbf{r}_i$ and particle $j$
at position $\mathbf{r}_j$.

The advantage of the Lennard-Jones potential is 
that it can be used to simulate a large variety of systems and scenarios. 
For example, each particle may represent an atom, a colloid, 
or a monomer of a polymer.
\cite{woodParker1957,kobAndersenPRL1994,guzmanDePablo2003,bennemann1998}
Depending on parameters such as the temperature, density, and 
shear stress the particles may form a gas, liquid, or solid
(crystal or glass).\cite{HansenVerlet1969,abramo2015,kobAndersenPRL1994}
The reason for this wide variety of applications is that the Lennard-Jones 
potential incorporates two major effective forces: a strong 
repulsive force for short distances and an attractive force 
for intermediate distances. 
The attractive term of the Lennard-Jones potential 
$\propto (\sigma/r)^6$ is the 
van der Waals interaction due to mutual polarization of two particles.
\cite{HansenMcDonald} 
The repulsive part $\propto (\sigma/r)^{12}$ is proportional to a 
power of $ (\sigma/r)^6$ and thus simplifies the computation of the force. 
Note that the Lennard-Jones potential is short-range.
For long-range interactions (gravitational and Coulomb)
more advanced techniques are necessary.\cite{allen90}
See Refs.~\onlinecite{allen90,rapaport98,HansenMcDonald} for an overview of further applications of the Lennard-Jones 
potential and other particle 
interactions and additional contributions to $\mathbf{F}$.

In this paper we illustrate how to simulate a glass forming system. 
We use the binary Kob-Anderson potential,
\cite{kobAndersenPRL1994,kobAndersenPRE51_1995,kobAndersenPRE52_1995} which has been developed as a model for the Ni$_{80}$P$_{20}$ alloy,\cite{kobAndersenPRE51_1995}
and has become one of the major models for 
studying supercooled liquids, glasses, and crystallization.
Examples are discussed in
Refs.~\onlinecite{kobAndersenPRL1994,kobAndersenPRE51_1995,kobAndersenPRE52_1995,KVLKobBinder_1996,hassani2016,schoenholz2016,ShrivastavPRE2016,makeevPriezjev2018,pedersenSchroederDyrePRL_2018} and references cited in Refs.~\onlinecite{kobAndersenPRL1994, makeevPriezjev2018,pedersenSchroederDyrePRL_2018}.
The Kob-Andersen model is an 80:20 mixture of particles of type $A$ and $B$.
The Lennard-Jones potential in Eq.~(\ref{eq:VLJ}) is 
modified by the dependence of $\epsilon$ and $\sigma$ 
on the particle type $\alpha,\beta \in \{A,B\}$ of particles $i$ and $j$. 
\begin{equation}
\label{eq:VKALJ}
 V_{ij}= V_{\alpha \beta}(r_{ij}) = 4\epsilon_{\alpha \beta} \left [ 
 \left ( \frac{\sigma_{\alpha \beta}}{r_{ij}} \right )^{12}
 - \left ( \frac{\sigma_{\alpha \beta}}{r_{ij}} \right )^{6} 
 \right ].
\end{equation}

We use units such that $\sigma_{AA}=1$ (length unit), 
$\epsilon_{AA}=1$ (energy unit), $m_A=1$ (mass unit), and 
$k_{\rm B}=1$ (the temperature unit is $\epsilon_{AA}/k_{\rm B}$).
The resulting time unit is $\sqrt{m_A \sigma_{AA}^2/\epsilon_{AA}}$.
With these units, the Kob-Andersen parameters are 
$\sigma_{AA}=1.0$, $\epsilon_{AA}=1.0$, 
$\sigma_{AB}=0.8$, $\epsilon_{AB}=1.5$, 
$\sigma_{BB}=0.88$, $\epsilon_{BB}=0.5$, and $m_A=m_B=1.0$. 
To save computer time, the potential 
is truncated and shifted at 
$r_{ij}=r^{\rm cut}_{\alpha \beta}=2.5 \,\sigma_{\alpha \beta}$
\begin{equation}
\label{eq:VshiftedKALJ}
 V_{ij}^{\rm cutoff}= 
\begin{cases}
 V_{\alpha \beta}(r_{ij}) - V_{\alpha \beta}(r^{\rm cut}_{\alpha \beta}) & \mbox{$r_{ij} < r^{\rm cut}_{\alpha \beta}$}\\
 0 & \mbox{(otherwise)}.
 \end{cases} 
\end{equation}
For the truncated and shifted KA-LJ system,
$V$ is given by Eq.~(\ref{eq:Vij}) by replacing
$V_{ij}$ with $V_{ij}^{\rm cutoff}$.

The force is given by
$
\mathbf{F}_i = - \nabla_i V$. The $x$-component of the force on particle $i$ 
is given by (see Problem~\ref{prob:Fix})
\begin{equation}
F_{i,x} = 48 \sum_{{\rm neighbors}\, j} \epsilon_{\alpha \beta}
             \left (\frac{\sigma_{\alpha \beta}^{12}}{r_{ij}^{14}} 
           -  0.5 \frac{\sigma_{\alpha \beta}^{6}}{r_{ij}^{8}}  \right ) (x_i - x_j).
\label{eq:KAFix}
\end{equation}
and similarly for $F_{i,y}$ and $F_{i,z}$.
The sum is only over particles $j$ for which $j \ne i$ and
$r_{ij} < 2.5 \, \sigma_{\alpha \beta}$.

\subsection{Periodic boundary conditions and the minimum image convention}
\label{sec:pbc}

\begin{figure}[t]
\begin{center}
\includegraphics[width=\columnwidth]{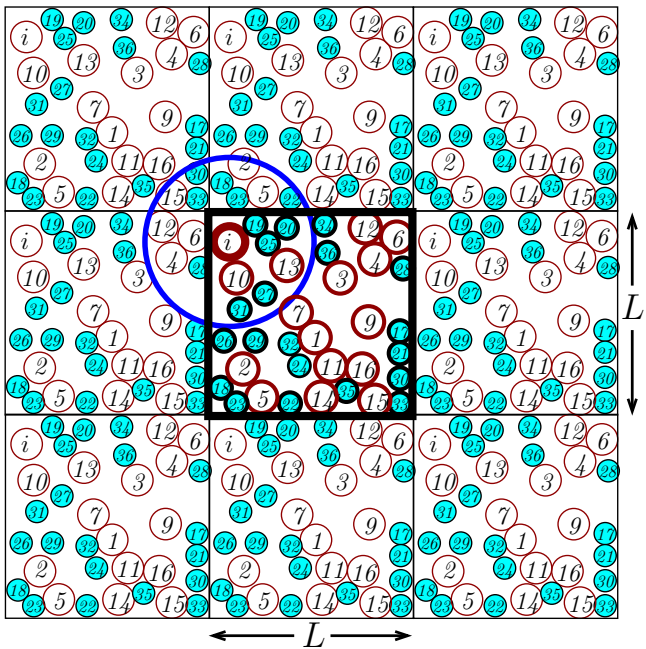}
\caption{Sketch of a binary two-dimensional system  
illustrating periodic boundary conditions and the minimum image 
convention. 
To identify the  neighbors $j$ of particle $i$, 
the position $\mathbf{r}_j$ of particle $j$  is chosen 
from the position of $j$ within the system 
(within thick frame) and the positions of $j$'s periodic images  
(within boxes framed with thin lines), such that
$r_{ij}$ is a minimum.  Neighbors of $i$ 
satisfy  $r_{ij} < r^{\rm cut}_{\alpha \beta}$ and are within 
the blue large circle.
} 
\label{fig:pbc}
\vspace{-0.3in}
\end{center}
\end{figure}

To determine the neighbors we need to specify 
the boundaries of the system. We will assume that 
the goal of the simulation 
is to model the structure and dynamics
of particles in a very large system ($N \approx 10^{23}$) 
far from the boundaries.
However, most molecular dynamics simulations contain on the order of $10^3$--$10^{6}$ particles. To 
minimize the effect of the boundaries, we use 
periodic boundary conditions 
as illustrated in Fig.~\ref{fig:pbc} for 
a two-dimensional system of linear dimension $L$. The system,
framed by thick lines, is assumed to be surrounded by
periodic images (framed by thin lines). For particle $i$
the neighboring particles within a distance 
$r^{\rm cut}_{\alpha \beta}$ 
are the particles inside the (blue) large circle. To determine the 
distance $r_{ij}$ between particles $i$ and $j$, 
we use the ``minimum image convention.''
For example, the distance between $i$ and particle $j=18$ 
would be $r_{ij} > r^{\rm cut}_{\alpha \beta}$ 
without using periodic images because particle 18  in the 
left bottom corner of the system is outside the (blue) large circle.  But with periodic images $r_{ij} < r^{\rm cut}_{\alpha \beta}$ because the nearest periodic image of particle 18 is above particle $i$ 
within the blue circle. For particles $i$ and $j=20$ we use the direct distance between the two
particles within the system (thick frame), 
because this distance is less than the distance 
to any of the periodic images of $j=20$.

\subsection{Numerical integration}
\label{sec:MDstep}

We next specify the core of a 
molecular dynamics program, that is, the numerical integration 
of the $d \times N$ coupled differential equations 
represented by  Eq.~(\ref{eq:Newton2ndlaw})  (see Fig.~\ref{fig:MDintro}). We will 
use the velocity Verlet algorithm 
\begin{align}
 \mathbf{r}_i \left ( t + \Delta t \right ) &= 
 \mathbf{r}_i(t) + \mathbf{v}_i(t) \Delta t
 + \frac{1}{2} \mathbf{a}_i(t) \left ( \Delta t \right )^2 
 \label{eq:velVerletPos} \\
 \mathbf{v}_i \left ( t + \Delta t \right ) &= 
 \mathbf{v}_i(t) 
 + \frac{1}{2} [ \mathbf{a}_i(t) +
 \mathbf{a}_i \left(t+\Delta t \right )
] \Delta t.
 \label{eq:velVerletVel}
\end{align}
The velocity Verlet algorithm is commonly used 
because it is energy drift free and second order 
in the velocity and third order in the 
position.\cite{gouldTobochnikChristian2007,allen90}
For other numerical integration techniques we refer readers to Refs.~\onlinecite{gouldTobochnikChristian2007,allen90,newman2013,numrecipes}.
Note that the velocity update 
in Eq.~(\ref{eq:velVerletVel}) is directly applicable 
only if $\mathbf{a}_i(t + \Delta t)$ does not depend on $\mathbf{v}_i(t+\Delta t)$;
that is, $a_i(t)$ depends only on the positions of the particles.

\begin{figure}[t]
\begin{center}
\includegraphics[width=\columnwidth]{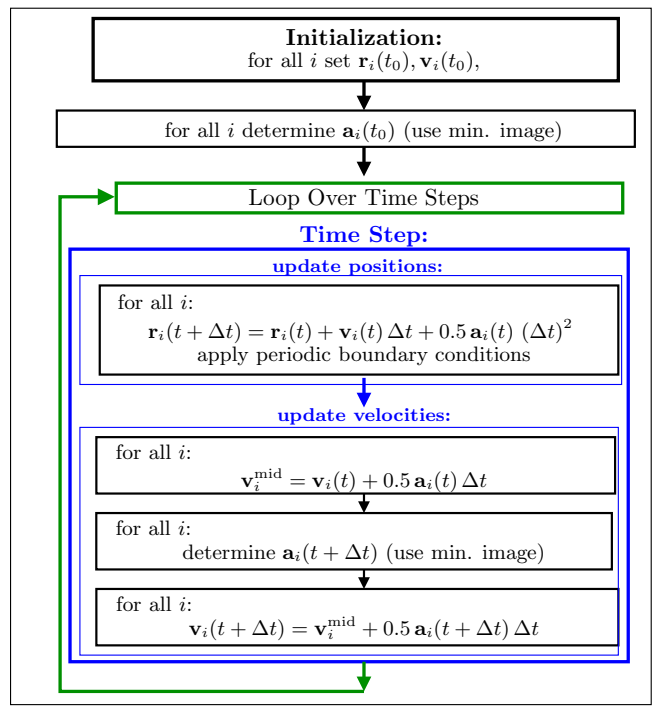}
\vspace{-0.2in}
\caption{Flow chart for  molecular dynamics with the velocity Verlet algorithm.} 
\label{fig:MDflowchartII}
\end{center}
\end{figure}

By using Eqs.~(\ref{eq:velVerletPos}) and (\ref{eq:velVerletVel}) we obtain
the flow chart (see Problem~\ref{prob:MDflowchart}) 
in Fig.~\ref{fig:MDflowchartII}. Most of the computational time is used to determine 
the accelerations.
Note that for each time step $\Delta t$ the accelerations need to be 
determined only once for $\mathbf{a}_i(t+\Delta t)$
(instead of twice for $\mathbf{a}_i(t)$ 
and $\mathbf{a}_i(t+\Delta t)$). Thus, 
only one array for the accelerations is needed, but one must have the correct 
order of updates within the time step.

\subsection{Temperature bath}
\label{sec:NVT}

So far we have used Newton's second law, Eq.~(\ref{eq:Newton2ndlaw}), 
and numerical integration to determine the dynamics, which corresponds 
to simulating a system at constant energy. We
also assumed that the number of particles and
the volume are constant, and therefore we have 
described the NVE or microcanonical ensemble.\cite{schroeder2000,blundells,gouldTobochnikStatMech}
In experiments the temperature $T$ and the pressure $P$ are controlled rather than $E$ and $V$. Many 
algorithms have been developed for 
NVT and NPT simulations, including generalizations 
which allow  box shapes to vary during the simulation. 
For an overview of these algorithms we recommend 
Refs.~\onlinecite{allen90,gouldTobochnikChristian2007,martynaMolPhys1996}.
In this section we focus on fixed NVT. We discuss in 
Sec.~\ref{sec:NVT_Boltzmann} 
an algorithm that also can be used to 
obtain the initial velocities 
and then discuss in Sec.~\ref{sec:NVT_NoseHoover}
the Nos\'{e}-Hoover algorithm. 
A generalization of the latter
is the default NVT algorithm in LAMMPS and is 
used in Sec.~\ref{sec:LAMMPSNVT}.

\subsubsection{Statistical temperature bath}
\label{sec:NVT_Boltzmann}

The canonical ensemble corresponds to a system that can exchange 
energy with a very large system at constant 
temperature $T$. In equilibrium the probability of a microstate 
$s$  is proportional to the Boltzmann factor 
\begin{equation}
 \label{eq:BoltzmannFactor}
 P(s) \propto
 \mathrm{e}^{- E(s)/k_{\rm B} T},
\end{equation}
where $k_{\rm B}$ is Boltzmann's 
constant.\cite{schroeder2000,blundells,gouldTobochnikStatMech}
Equation~(\ref{eq:BoltzmannFactor}) applies to any 
system. The microstate $s$ is specified by
the position and velocity of each particle,
$\{\mathbf{r}_i,\mathbf{v}_i\}$, 
and the system energy is
\begin{equation}
 \label{eq:EsysNpart}
 E\left(\{\mathbf{r}_i,\mathbf{v}_i\}\right) = 
 \frac{1}{2} \sum \limits_{i=1}^N m_i \mathbf{v}_i^2 
 + V(\{\mathbf{r}_i\}).
\end{equation}
From Eqs.~(\ref{eq:BoltzmannFactor}) and (\ref{eq:EsysNpart}) 
it follows that the probability distribution for 
the $x$-component of the velocity of particle $i$
is given by the Maxwell-Boltzmann distribution 
\begin{equation}
 \label{eq:vxocitydistribution}
 P\left(v_{i,x}\right) = 
 \frac{1}{\sqrt{2 \pi} \sigma_i} 
 \mathrm{e}^{- v_{i,x}^2/2 \sigma_i^2},
\end{equation}
with the standard deviation
\begin{equation}
 \label{eq:sigmavx}
 \sigma_i = \sqrt{\frac{k_{\rm B} T}{m_i}}.
\end{equation}
The probability distributions for the $y$- and $z$-components
of the velocity are obtained by replacing in 
Eq.~(\ref{eq:vxocitydistribution})
$v_{i,x}$ by $v_{i,y}$ and $v_{i,z}$, respectively.
\cite{schroeder2000,blundells,gouldTobochnikStatMech}
It is straightforward to show that (see Problem~\ref{prob:EkinTrelation})
\begin{equation}
 \label{eq:equip3N}
 \left \langle E_{\rm kin} \right \rangle = 
 \left \langle \sum \limits_{i=1}^{N} \frac{1}{2} m_i v_i^2 \right \rangle 
 = \frac{3 N}{2} k_{\rm B} T .
\end{equation}

To achieve simulations at fixed NVT Andersen\cite{andersen1980}
incorporated particle collisions with a temperature bath by choosing the particle
velocities from the Maxwell-Boltzmann 
distribution.\cite{allen90,gouldTobochnikChristian2007} We will use 
a slight modification to the Andersen 
algorithm introduced by Andrea et al.\cite{andrea} 
At periodic intervals 
(approximately every $50$ time steps) 
all velocities are newly assigned by giving each particle a velocity component $v_{\mu}$
($\mu \in \{x,y,z\}$) chosen from 
the Maxwell-Boltzmann distribution in 
Eqs.~(\ref{eq:vxocitydistribution}) 
and (\ref{eq:sigmavx}).\cite{allen90}

\begin{figure}[t]
\begin{center}
\includegraphics[width=\columnwidth]{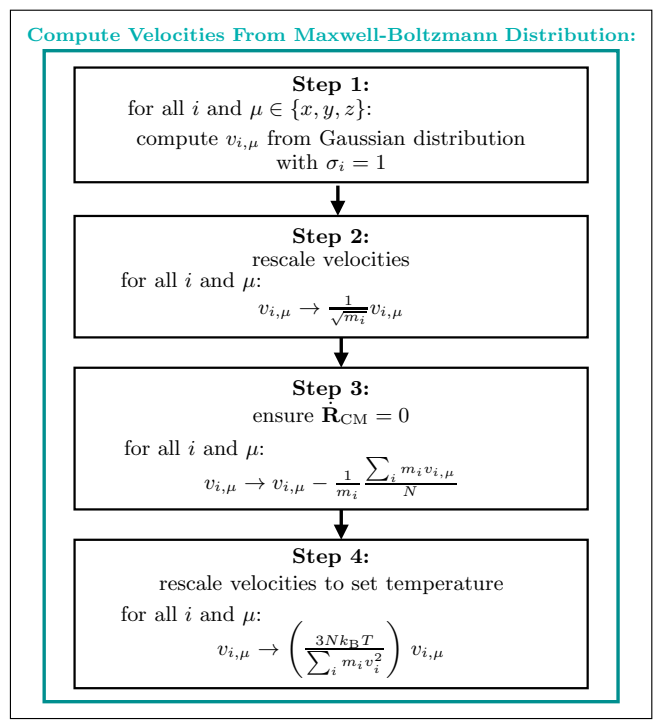}
\vspace{-0.2in}
\caption{Flow chart for computing the velocities from the 
Maxwell-Boltzmann distribution.} 
\label{fig:VelBoltzmannFlowChart}
\end{center}
\end{figure}

Figure~\ref{fig:VelBoltzmannFlowChart} shows the flow chart 
for creating a Maxwell-Boltzmann distribution for the velocities of the particles.
Step 1 can be done in any programming language with either 
already defined functions (see Sec.~\ref{sec:pythonInitializationVel}
for Python and Sec.~\ref{sec:LAMMPNVE} for LAMMPS)
or with functions (or subroutines), from for example, Ref.~\onlinecite{numrecipes}. Step 2 can be skipped
when $m_i$ is the same for all particles. Step 3 ensures that the center of mass 
of the system does not drift, and step 4 rescales all the
velocities to achieve the desired temperature.

The computer code illustrated by 
the flow chart of Fig.~\ref{fig:VelBoltzmannFlowChart} is 
inserted into the code described by the flow chart of Fig.~\ref{fig:MDflowchartII}
with a conditional statement (e.g., \verb+if+) 
after the (blue) 
``Time Step'' 
box and within 
the (green) ``Loop Over Time Steps.'' We can also apply steps 1---4 
to set the initial velocities $\{\mathbf{v}_i(t_0)\}$
as part of the ``Initialization'' box in 
Fig.~\ref{fig:MDflowchartII}.

\subsubsection{Nos\'{e}-Hoover algorithm}
\label{sec:NVT_NoseHoover}

Another way to implement a constant temperature bath, which is
used in LAMMPS,\cite{LAMMPSfixnvtp}
is given by the Nos\'{e}-Hoover 
style
algorithm. The key concept for most of the advanced 
algorithms is that we no longer use Newton's second law 
for the equations of motion, but instead use an ``extended 
system'' with additional parameters. For the
Nos\'{e}-Hoover algorithm\cite{nosehoover} 
Eq.~(\ref{eq:Newton2ndlaw}) is 
replaced by 
\begin{align}
 \ddot{\mathbf r}_i & = \frac{{\mathbf F}_i}{m_i} - \xi \, \dot{\mathbf r}_i ,
 \label{eq:nosehoover_rddot}\\
 \frac{{\rm d}^2 \ln s}{{\rm d}t^2} = \dot{\xi} 
 & = \frac{1}{Q} 
 \left ( \sum \limits_{i=1}^{N} m_i \dot{{\mathbf r}}_i^2 - dN k_{\rm B} T
 \right ), \label{eq:nosehoover_xidot}
\end{align}
where $d$ is the spatial dimension. 
The idea is to introduce a fictitious dynamical variable $\xi$ that plays the role of a friction which changes the acceleration until the temperature equals the desired value.  The parameter $Q$  is the mass of the temperature bath.
Equations~(\ref{eq:nosehoover_rddot}) and (\ref{eq:nosehoover_xidot})
follow from generalizations of Hamiltonian
mechanics\cite{Taylor} and 
can be written as 
first-order differential
equations for $\dot{\mathbf{r}}_i$, $\dot{\mathbf{p}}_i$,
and $\dot{\xi}$.
\cite{nosehoover,Frenkel2002,nose1984,branka,martynaJCP1994,martynaMolPhys1996,tuckerman2001} 
A simplified derivation of these equations is given in Appendix~\ref{sec:derivationNoseHoover} for
readers who are familiar with Hamiltonian mechanics.
 
A generalization of
Eqs.~\eqref{eq:nosehoover_rddot} and (\ref{eq:nosehoover_xidot}) 
is the Nos\'{e}-Hoover chain method, which includes variables for several
temperature baths, corresponding to more accurate dynamics
in cases with more constraints than the example presented in this paper. 
\cite{shinoda2004,martynaKleinTuckermanJCP1992,martynaMolPhys1996,martynaJCP1994,Frenkel2002}

Given the equations of motion, our task is to 
convert them to appropriate difference equations so that we can use numerical 
integration. We would like to use the velocity Verlet algorithm in
Eqs.~(\ref{eq:velVerletPos}) and (\ref{eq:velVerletVel}). However, 
Eq.~(\ref{eq:nosehoover_rddot}) 
for $\mathbf{a}_i$ depends on the velocity $\dot{\mathbf{r}}_i$, which 
means that the right side of Eq.~(\ref{eq:velVerletVel}) also 
depends on $\mathbf{v}_i(t+\Delta t)$. Fox and Andersen
\cite{fox} suggested a velocity-Verlet numerical integration technique
that can be applied when the equations of motion are 
of the form
\begin{align}
 \ddot{\mathbf{x}}(t) & = 
 f[\mathbf{x}(t),\dot{\mathbf{x}}(t),\mathbf{y}(t),\dot{\mathbf{y}}(t)] \\
 \ddot{\mathbf{y}}(t) & = g[\mathbf{x}(t),\dot{\mathbf{x}}(t),\mathbf{y}(t)] .
\end{align}
The Fox-Andersen integration technique and 
its application to the Nos\'{e}-Hoover equations of motion,
Eqs.~(\ref{eq:nosehoover_rddot}) and (\ref{eq:nosehoover_xidot}), are discussed in Appendix~\ref{sec:FoxAndersonNoseHoover}. 
The resulting update rules are given in
Eqs.~(\ref{eq:Noseupdate_ri})--(\ref{eq:Noseupdate_xiapprox}), 
and (\ref{eq:ridotKVLRoman}). 
For more advanced integration techniques see Ref.~\onlinecite{tapiasJPC2016}.

\subsection{Initialization of positions and velocities}
\label{sec:initialization}

As shown in Figs.~\ref{fig:MDintro} and
~\ref{fig:MDflowchartII}, a molecular dynamics simulation starts 
with the initialization of every particle's position and 
velocity. (For  more complicated systems, further variables such as angular velocities need to be initialized.) As noted in Ref.~\onlinecite{gouldTobochnikChristian2007},  Sec.~8.6,  
``An appropriate choice of the initial conditions is more difficult 
than might first appear.'' We therefore discuss a few
options in detail. The most common options for the initialization of the particle 
positions include (1) using the positions resulting from 
 a previous simulation of the same system, (2) choosing uniformly distributed positions at random, or (3) starting with positions on a lattice. For simplicity, the latter may be on a cubic lattice and/or a crystalline structure.

The advantage of the first option is that the configuration might correspond to
a well equilibrated system at the desired parameters and the updates do not need extra precautions
as in options~2 and 3. Even if the available configuration is not 
exactly for the desired parameters, it might be appropriate to adjust 
the configuration (for example, to rescale all positions to obtain the desired
density) to avoid the disadvantages of the other options. 

Option~2 has the advantage that it provides at least some 
starting configurations if the other options are not possible.
The disadvantage is that if a few of the particles are too close to each other,  very large forces will result.
The large
forces can lead to runaway positions and velocities and thus additional steps must be taken. One idea
is to  rearrange the particle positions corresponding
to the local potential minimum. This rearrangement can be
achieved with a minimization program and/or with successive 
short simulation runs.
We can start with a very small time step $\Delta t$ and a very low 
temperature $T$ and successively increase both $\Delta t$ and $T$.

The advantage of option~3 is that very large forces 
are avoided. However, the lattice structure is not desirable 
for  studying  systems without long-range order, for example, 
supercooled liquids and glasses. 
A sequence of sufficiently long simulation runs 
may overcome this problem. 
For example,
the system might first successively be heated and then
quenched to the desired temperature. 
If the system has more than one particle 
type, the mixing of the particle types should be ensured.
(In Sec.~\ref{sec:pythonInitializationPos}
we provide an example where $A$ and 
$B$ particles are randomly swapped.)

Common options for the initialization of the velocities include (1) using the velocities from a previous simulation of the same system, (2) computing the velocities from the Maxwell-Boltzmann distribution corresponding to the desired temperature, and (3) setting all velocities to zero. If previous simulation configurations are available at the desired parameters, that option is always the best.
Option~2 is the most common initialization of velocities because
it corresponds to velocities of a well equilibrated system (see 
Sec.~\ref{sec:NVT_Boltzmann}). Option~3 is an option for $T=0$ simulations, which we will not
discuss further.

\section{Implementation of MD simulations with Python}
\label{sec:MDPython}

In this section, we assume that readers know how 
to write  basic Python programs. For the newcomer
without Python experience, we recommend the first few chapters 
of Ref.~\onlinecite{newman2013}, which are available online,
\cite{newmanOnline} and/or  other online resources. 
(Reference~\onlinecite{newmanOnline} includes external links.)

\subsection{Initialization of positions}
\label{sec:pythonInitializationPos}
 
At the beginning of the  simulation the initial 
configuration needs to be set. We use arrays for $\mathbf{r}$
and $\mathbf{v}$ of size $N=N_A+N_B$. For 
$N_A=800$ and $N_B=200$ the Python commands are 
\begin{verbatim}
import numpy as np
global Na
global Nb
global N
Na=800
Nb=200
N=Na+Nb
x = np.zeros(N,float)
y = np.zeros(N,float)
z = np.zeros(N,float)
\end{verbatim}

If the positions are available, 
we read them from a file. We assume the filename 
\verb+initpos+ contains $N$ lines each with 
three columns for $r_{i,x}$, $r_{i,y}$, and $r_{i,z}$ and
assign the positions using the statement
\begin{widetext}
\begin{verbatim}
x,y,z = sp.loadtxt('initpos',dtype='float',unpack=True)
\end{verbatim}
 
To choose positions at random and to ensure reproducible  results we set the seed once 
at the beginning of the program
\begin{verbatim}
import scipy as sp
sp.random.seed(15)
\end{verbatim}
For a system of linear dimension $L= 9.4$ we set the positions with
\begin{verbatim}
L = 9.4
x,y,z = sp.random.uniform(low=0.0,high=L,size=(3,N))
\end{verbatim}
\end{widetext}

For simplicity, we use the simple cubic lattice to place  the particles 
on lattice sites (see Problem~\ref{prob:poslattice})
as in the following:
\begin{verbatim}
nsitesx = int(round(pow(N,(1.0/3.0))))
dsitesx = L / float(nsitesx)
for ni in range(nsitesx):
  tmpz = (0.5 + ni)*dsitesx
  for nj in range(nsitesx):
    tmpy = (0.5 + nj)*dsitesx
    for nk in range(nsitesx):
      tmpx = (0.5 + nk)*dsitesx
      i=nk+nj*nsitesx+ni*(nsitesx**2)
      x[i]=tmpx
      y[i]=tmpy
      z[i]=tmpz
\end{verbatim}

The arrays \verb+x+, \verb+y+, \verb+z+ use the 
indices $0,1, \ldots, N_A-1$ to store the positions of the $A$ particles. Because the 
lattice positions are assigned successively to the lattice, the
code places all $A$ particles on one side and all $B$ 
particles on the other side. This arrangement is not what is intended  for 
a glassy or supercooled  system. Therefore, in addition to
assigning lattice sites, we next swap each $B$ particle's position 
with a randomly chosen $A$ particle's position:
\begin{verbatim}
for i in range(Na,N):
   j = sp.random.randint(Na)
   x[i],x[j] = x[j],x[i]
   y[i],y[j] = y[j],y[i]
   z[i],z[j] = z[j],z[i]
\end{verbatim}

\subsection{Plotting and visualization tools}
\label{sec:pythonPlotting}

Long programs  should be divided   into
many smaller tasks and  each task tested.  
Before we continue with the implementation of molecular dynamics,
we  use a few printing and plotting tools to check if
the program is working as expected.

We can save the positions into a file with a name such as \verb+initposcheck+
\begin{widetext}
\begin{verbatim}
sp.savetxt('initposcheck', (sp.transpose(sp.vstack((x,y,z)))))
\end{verbatim}
\end{widetext}
and then check  the numbers in the  file or  look at the positions visually. To plot the positions 
we can use either plotting commands such as \verb+gnuplot+, 
\verb+xmgrace+, or Python-plotting tools. For simplicity,
we use the latter. The Python commands for making
a two-dimensional scatter plot of $r_{i,z}$ and $r_{i,x}$
which distinguishes $A$ and $B$ particles by color and size are
\begin{verbatim}
import matplotlib as mpl
import matplotlib.pyplot as plt 
plt.figure()
plt.scatter(x[:Na],z[:Na],s=150,color='blue')
plt.scatter(x[Na:],z[Na:],s=70,color='red')
plt.xlim(0,L)
plt.xlabel('$x$')
plt.ylabel('$z$')
plt.show()
\end{verbatim}

Figure~\ref{fig:posfigures}(a) shows the resulting scatter plot 
for the  case where the initial positions are on a lattice.
To make a three-dimensional scatter plot we use at the beginning of 
the  program the same \verb+import+ commands and 
add the line 
\begin{verbatim}
from mpl_toolkits.mplot3d import Axes3D 
\end{verbatim}
and then
\begin{widetext}
\begin{verbatim}
fig3d = plt.figure()
fax = fig3d.add_subplot(111, projection='3d')
fax.scatter(x[:Na],y[:Na],z[:Na], marker="o",s=150,facecolor='blue')
fax.scatter(x[Na:],y[Na:],z[Na:], marker="o",s=70,facecolor='red')
fax.set_xlabel('$x$')
fax.set_ylabel('$y$')
fax.set_zlabel('$z$')
plt.show()
\end{verbatim}
The resulting figure is shown in Fig.~\ref{fig:posfigures}(b).
Use  the right mouse button to  zoom in and out and  the left
mouse button  to rotate the figure. This three-dimensional scatter plot is useful for a quick and easy check.

\begin{figure}[t]
\begin{center}
\includegraphics[width=0.9 \textwidth]{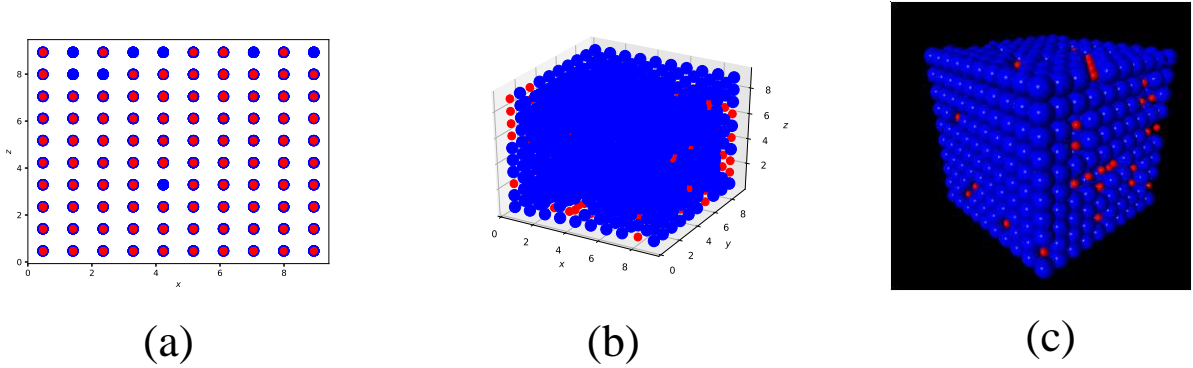}
\caption{Visualizations of the particle positions, which were assigned initially to a cubic lattice.} 
\label{fig:posfigures}
\end{center}
\end{figure}

For a fancier three-dimensional visualization of the particles, the powerful
package \verb+VPython+ is useful. 
\begin{verbatim}
from vpython import *
for i in range(N):
  tx=x[i]
  ty=y[i]
  tz=z[i]
  if i < Na :
    sphere(pos=vector(tx,ty,tz),radius=0.5,color=color.blue)
  else:
    sphere(pos=vector(tx,ty,tz),radius=0.2,color=color.red)
\end{verbatim}
Use the right mouse button to  rotate the plot and  the middle 
mouse button to  zoom in and out [see Fig.~\ref{fig:posfigures}(c)]. 
More information on plotting tools is available at 
Ref.~\onlinecite{newmanOnline} 
and their external links.
The Python program and the initial configuration file for
this section are in the files
\verb+KALJ_initpos.py+ and \verb+initpos+ available at Ref.~\onlinecite{supplement}.

\subsection{Initialization of velocities}
\label{sec:pythonInitializationVel}

If the positions and velocities are already available, they can be read from a file with six columns, similar to our earlier example. As described in Sec.~\ref{sec:NVT_Boltzmann}
a temperature bath can be achieved by periodically 
resetting all velocities from the Maxwell-Boltzmann distribution at the 
desired temperature. We therefore define a function 
for this task.
\begin{verbatim}
def maxwellboltzmannvel(temp):
  global vx
  global vy
  global vz
  nopart=len(vx)
  sigma=np.sqrt(temp) #sqrt(kT/m)
  vx=np.random.normal(0.0,sigma,nopart)
  vy=np.random.normal(0.0,sigma,nopart)
  vz=np.random.normal(0.0,sigma,nopart)
#  make sure that center of mass does not drift
  vx -= sum(vx)/float(nopart)
  vy -= sum(vy)/float(nopart)
  vz -= sum(vz)/float(nopart)
#  make sure that temperature is exactly wanted temperature
  scalefactor = np.sqrt(3.0*temp*nopart/sum(vx*vx+vy*vy+vz*vz))
  vx *= scalefactor
  vy *= scalefactor
  vz *= scalefactor
\end{verbatim}
This function is used in the example using the Python command 
{\tt maxwellboltzmannvel(0.2)}  with $T=0.2$.
To check the resulting velocities we save them in a file, 
plot them, and/or visualize them in \verb+VPython+ with
\begin{verbatim}
arrow(pos=vector(tx,ty,tz), axis=vector(tvx,tvy,tvz),color=color.green)
\end{verbatim}
where \verb+tx,ty,tz+ correspond to the positions 
and \verb+tvx,tvy,tvz+ correspond to the velocities. The implementation for this section is in
\verb+KALJ_initposvel.py+.\cite{supplement}
\end{widetext}

\subsection{Accelerations}
\label{sec:pythonAccelerations}
 
As shown in Fig.~\ref{fig:MDflowchartII}, the
accelerations $\mathbf{a}_i$ are determined after the 
initialization and at each  time step. 
A user-defined function for  determining the accelerations
is therefore recommended.(see Problem~\ref{prob:accfct}).
For the KA-LJ system, $m_i=1$ and $F_{i,x}$ as given in
Eq.~(\ref{eq:KAFix}) (similarly $F_{i,y}$ and $F_{i,z}$).
 We store the accelerations 
in arrays \verb+ax, ay, az+. To determine $\mathbf{a}_i$ for all 
$i=1,\ldots, N$ (in Python particle index \verb+i=0,+$\ldots$\verb+,N-1+),
we need a loop over $i$ and  implement the sum over neighbors 
using an inner loop over $j$. From Newton's third law,
$\mathbf{F}_{ij}=-\mathbf{F}_{ji}$, we can reduce this double sum 
by a half.
\begin{verbatim}
def acceleration(x,y,z):
  ax=sp.zeros(N)
  ay=sp.zeros(N)
  az=sp.zeros(N)
  for i in range(N-1):
    ...
    for j in range(i+1,N):
      ...
      ax[i] += ...
      ax[j] -= ...
      ...
\end{verbatim}
To determine $r_{ij}^2$ we use \verb+rijto2+,
\begin{verbatim}
def acceleration(x,y,z):
  ax=sp.zeros(N)
  ...
  for i in range(N-1):
    xi=x[i]
    yi=y[i]
    zi=z[i]
    for j in range(i+1,N):
      xij=xi-x[j]
      yij=yi-y[j]
      zij=zi-z[j]
      ...
      rijto2 = xij*xij + yij*yij + zij*zij
      ...
\end{verbatim}
In addition, we need to implement the minimum image convention 
as described in Sec.~\ref{sec:pbc}.
\begin{verbatim}
def acceleration(x,y,z):
  global L
  global Ldiv2
   ...
      xij=xi-x[j]
      yij=yi-y[j]
      zij=zi-z[j]
      # minimum image convention
      if xij > Ldiv2: xij -= L
      elif xij < - Ldiv2: xij  += L
      if yij > Ldiv2: yij -= L
      elif yij < - Ldiv2: yij  += L
      if zij > Ldiv2: zij -= L
      elif zij < - Ldiv2: zij  += L
      rijto2 = xij*xij + yij*yij + zij*zij
      ...
\end{verbatim}
Here \verb+Ldiv2+ is $L/2.0$, and we assume  the
particle positions satisfy 
$0 < x,y,z < L$, which means that \verb+x+,
\verb+y+, \verb+z+ are updated with periodic boundary conditions
to stay within the central simulation box.
Because  the sum is only over neighbors within $2.5\sigma$ of a given particle,  
we add an if statement:
\begin{widetext}
\begin{verbatim}
def acceleration(x,y,z):
  ...
     rijto2 = xij*xij + yij*yij + zij*zij
     if(rijto2 < rcutto2):
       onedivrijto2 = 1.0/rijto2
       fmagtmp= eps*(sigmato12*onedivrijto2**7 - 0.5*sigmato6*onedivrijto2**4)
       ax[i] += fmagtmp*xij
       ax[j] -= fmagtmp*xij
       ...
  return 48*ax,48*ay,48*az
\end{verbatim}
\end{widetext}
We avoided the additional costly computational determination
of $r_{ij}=\sqrt{r_{ij}^2}$, and the factor \verb+48+ was
multiplied only once via matrix multiplication. This acceleration function is called in the main program with the statement
\begin{verbatim}
ax,ay,az = acceleration(x,y,z)
\end{verbatim}

The variables $\mbox{\tt rcutto2}=(r^{\rm cut})^2$,
$\mbox{\tt eps}=\epsilon$, $\mbox{\tt sigmato12}=\sigma^{12}$, and 
$\mbox{\tt sigmato6}=\sigma^{6}$ are 
particle type dependent for the binary Kob-Andersen model. We  include this dependence 
with conditional statements
\begin{verbatim}
...
  for i in range(N-1):
    ...
    for j in range(i+1,N):
      ...
      if i < Na:
        if j < Na:   #AA
           rcutto2 = rcutAAto2
           sigmato12 = sigmaAAto12
           sigmato6 = sigmaAAto6
           eps = epsAA
        else:            #AB
           rcutto2 = rcutABto2
           sigmato12 = sigmaABto12
           sigmato6 = sigmaABto6
           eps = epsAB
      else:              #BB
\end{verbatim}
These conditional statements cost computer time  and can 
be avoided by replacing the {\tt i}, {\tt j} loops with three separate {\tt i}, {\tt j} loops.
\begin{widetext}
\begin{verbatim}
# AA interactions
  for i in range(Na-1):
    ...
    for j in range(i+1,Na):
      ...
      rijto2 = xij*xij + yij*yij + zij*zij
      if(rijto2 < rcutAAto2):
        onedivrijto2 = 1.0/rijto2
        fmagtmp= epsAA*(sigmaAAto12*onedivrijto2**7 - 0.5*sigmaAAto6*onedivrijto2**4)
        ...
# AB interactions
  for i in range(Na):
    ...
    for j in range(Na,N):
      ...
      rijto2 = xij*xij + yij*yij + zij*zij
      if(rijto2 < rcutABto2):
        onedivrijto2 = 1.0/rijto2
        fmagtmp= epsAB*(sigmaABto12*onedivrijto2**7 - 0.5*sigmaABto6*onedivrijto2**4)
        ...
# BB interactions
  for i in range(Na,N-1):
    ...
    for j in range(i+1,N):
      ...
  return 48*ax,48*ay,48*az
\end{verbatim}
\end{widetext}

\subsection{NVE molecular dynamics simulation}
\label{sec:pythonNVEMD}

We are now  equipped to implement a NVE molecular dynamics 
simulation (see Problem~\ref{prob:NVEMD}).
We follow the flow chart of Fig.~\ref{fig:MDflowchartII} 
and  after the initialization of $\left\{\mathbf{r}_i\right\}$, 
$\left\{\mathbf{v}_i\right\}$, and $\left\{\mathbf{a}_i\right\}$, add
a loop over time steps and  update the 
positions and velocities within this loop. To update $\mathbf{r}_i$ 
and $\mathbf{v}_i$ for all $i$, we use matrix operations instead of 
{\tt for} loops, because  matrix operations are computationally
faster in Python. We need to ensure periodic boundary conditions,
which we implement 
assuming that we start with $0 < x_i, y_i, z_i \le L$ and that 
during each  time step no particle moves further than $L$.
\begin{verbatim}
for tstep in range(1,nMD+1):
  # update positions
  x += vx*Deltat + 0.5*ax*Deltatto2
  y += vy*Deltat + 0.5*ay*Deltatto2
  z += vz*Deltat + 0.5*az*Deltatto2
  # periodic boundary conditions:
  for i in range(N):
    if x[i] > L: x[i] -= L
    elif x[i] <= 0:  x[i] += L
    if y[i] > L: y[i] -= L
    elif y[i] <= 0:  y[i] += L
    if z[i] > L: z[i] -= L
    elif z[i] <= 0:  z[i] += L
  # update velocities
  vx += 0.5*ax*Deltat
  vy += 0.5*ay*Deltat
  vz += 0.5*az*Deltat
  ax,ay,az = acceleration(x,y,z)
  vx += 0.5*ax*Deltat
  vy += 0.5*ay*Deltat
  vz += 0.5*az*Deltat
\end{verbatim}
Here ${\tt{nMD}}=n_{\rm MD}$ is the number of time steps,
${\tt Deltat}=\Delta t$, and ${\tt Deltatto2}=(\Delta t)^2$.

The most time consuming part of the time loop 
is the determination of $\left\{\mathbf{a}_i\right \}$, because 
it includes the double loop over $i$ and $j$. In more optimized 
MD programs, the loop over $j$ would be significantly sped up by
looping only over neighbors of particle $i$ (instead of over all particles $j$)
via a   neighbor list.\cite{allen90}
However, for our purpose of becoming familiar with  
MD,
we will do  without a neighbor list.

\subsection{Python: More analysis and visualization tools}
\label{sec:pythonPlottingII}

\begin{figure}[h]
\begin{center}
\includegraphics[width=\columnwidth]{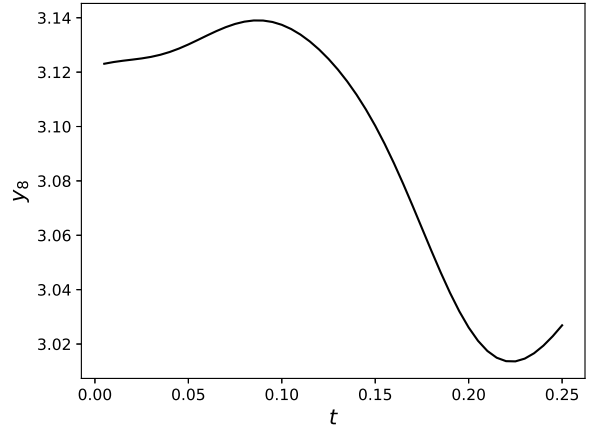}
\vspace{-0.1in}
\caption{The $y$-component of   particle 8 
as a function of time $t$ 
(in LJ time units $\sqrt{m_A \sigma_{AA}^2/\epsilon_{AA}}$)
for the Kob-Andersen
model of $N=1000$ particles
starting with the initial configuration {\tt initposvel}, and 
for $50$ time steps 
with $\Delta t=0.005$.
(The largest time is $50 \times 0.005 = 0.25$.)
}
\label{fig:y8oft}
\end{center}
\end{figure}

\begin{figure}[t]
\begin{center}
\includegraphics[width=\columnwidth]{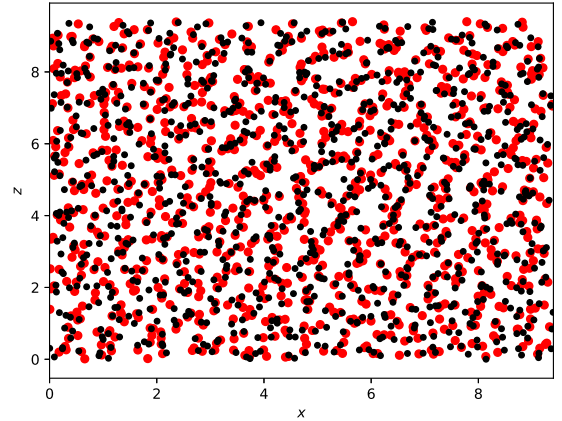}
\caption{$z(x)$  for the initial configuration in black small circles 
and for the 
final configuration 
at $t=0.25$ (in LJ time units $\sqrt{m_A \sigma_{AA}^2/\epsilon_{AA}}$)
in red large circles.}
\label{fig:initfinalzofx}
\end{center}
\end{figure}

As a check of the program, we may either plot the trajectories
of 
a specified particle (see Fig.~\ref{fig:y8oft}) or   make a
scatter plot of the initial and final configuration 
as shown in Fig.~\ref{fig:initfinalzofx} for $z(x)$ after 
$n_{\rm MD}=50$ time steps with $\Delta t = 0.005$.

For Fig.~\ref{fig:y8oft} we used the plot tools
as described in Sec.~\ref{sec:pythonPlotting}. To plot the
$y$-component of  particle 8,  
an array for these values was defined before the time loop 
\begin{verbatim}
yiplotarray = np.zeros(nMD,float)
\end{verbatim}
This array was updated within the time loop after the time step
\begin{verbatim}
yiplotarray[tstep-1]=y[7]
\end{verbatim}
Note that   particle 8 corresponds to index {\tt 7}.
After the time loop the following plotting commands are used:
\begin{verbatim}
tarray = np.arange(Deltat,(nMD+1)*Deltat,Deltat)
plt.rcParams['xtick.labelsize']=11
plt.rcParams['ytick.labelsize']=11
plt.figure()
plt.plot(tarray,yiplotarray,color='blue')
plt.xlabel('$t$',fontsize=15)
plt.ylabel('$y_8$',fontsize=15)
plt.show()
\end{verbatim}

To make Fig.~\ref{fig:initfinalzofx}
we stored the initial configuration using 
\begin{verbatim}
x0 = np.copy(x)
y0 = np.copy(y)
z0 = np.copy(z)
\end{verbatim}
and used {\tt plt.scatter} as described 
in Sec.~\ref{sec:pythonPlotting}. 

We can also make an animation using {\tt VPython}.\cite{newmanOnline} Before the time loop we create  spheres (particles at their positions)
and arrows (velocities) as
\begin{widetext}
\begin{verbatim}
s = np.empty(N,sphere)
ar = np.empty(N,arrow)
for i in range(N):
  if i < Na :
    s[i] = sphere(pos=vector(x[i],y[i],z[i]),radius=0.5,color=color.blue)
  else:
    s[i] = sphere(pos=vector(x[i],y[i],z[i]),radius=0.2,color=color.red)
  ar[i]=arrow(pos=vector(x[i],y[i],z[i]),axis=vector(vx[i],vy[i],vz[i]),
          color=color.green)  
\end{verbatim}
\end{widetext}
Within the time loop we update the spheres and arrows as
\begin{verbatim}
  rate(30)
  for i in range(N):
    s[i].pos = vector(x[i],y[i],z[i])
    ar[i].pos = vector(x[i],y[i],z[i])
    ar[i].axis = vector(vx[i],vy[i],vz[i])
\end{verbatim}

We can also plot 
the kinetic energy per particle $E_{\rm kin}/N$, potential energy 
per particle $V/N$, and the total energy per particle 
$E_{\rm tot}/N=\left(E_{\rm kin}+V\right)/N$ as a function of time.  
To include the minimum image 
convention and the cases $AA$, $AB$, and $BB$, a user-defined 
function can be written similar to
\verb+acceleration+ of Sec.~\ref{sec:pythonAccelerations}. The Python program for this section, {\tt KALJ\_nve.py}, including the accelerations in Sec.~\ref{sec:pythonAccelerations}, is available in Ref.~\onlinecite{supplement}.

\begin{figure}[h]
\begin{center}
\includegraphics[width=\columnwidth]{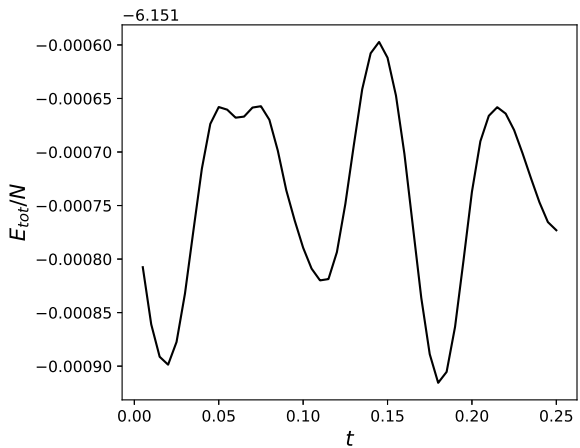}
\caption{The total energy per particle 
$E_{\rm tot}/N=\left(E_{\rm kin}+V\right)/N$ 
as a function of time 
(in LJ units) 
for a NVE  simulation 
with $n_{\rm MD}=50$ and with $\Delta t=0.005$.
The value $-6.161$ on
the top left indicates that the tick labels on the vertical axis are
$E_{\rm tot}/N+6.151$.
The initial positions and velocities were read in from
the file {\tt initposvel} ($T=0.5$).
}
\label{fig:Etotoft}
\end{center}
\end{figure}

As shown in Fig.~\ref{fig:Etotoft}, $E_{\rm tot}/N$ exhibits very small variations
about a constant as expected for the NVE  simulation. 

\subsection{NVT molecular dynamics simulation}
\label{sec:pythonNVTMD}

We now
implement a
statistical temperature bath as described in 
Sec.~\ref{sec:NVT_Boltzmann}, which will allow  
readers to do simulations   
at desired temperatures (see Problem~\ref{prob:NVTstochastic}).

\begin{figure}[t]
\begin{center}
\includegraphics[width=\columnwidth]{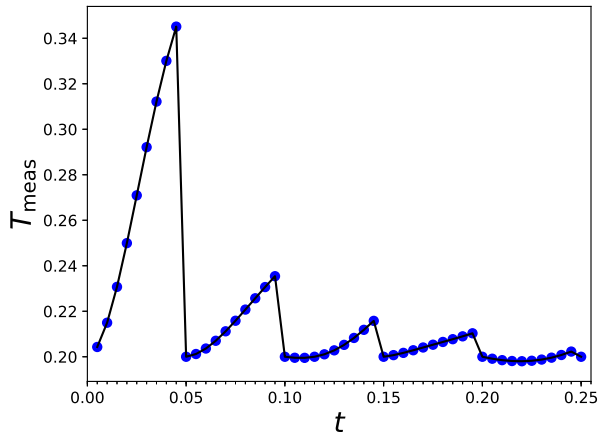}
\caption{The time-dependence of the temperature  
for fixed NVT   with $T=0.2$, ${\tt nstepBoltz}=10$,
and $n_{\rm MD}=50$   with $\Delta t=0.005$.
}
\label{fig:Toft}
\end{center}
\end{figure}

We need to update all the velocities using the Maxwell-Boltzmann distribution specified in 
Eqs.~(\ref{eq:vxocitydistribution}) and (\ref{eq:sigmavx}).
Because we already have implemented the Maxwell-Boltzmann 
distribution in Sec.~\ref{sec:pythonInitializationVel}
with the user-defined function 
\verb+maxwellboltzmannvel+, we can 
implement the temperature bath with only two lines 
in the time loop after the time step. 
For the case of computing new velocities every
$\mbox{\tt nstepBoltz}$ time steps: 
\begin{verbatim}
  if (tstep % nstepBoltz) == 0 :
    maxwellboltzmannvel(temperature)
\end{verbatim}

You can test your program by plotting the 
temperature as a function of time. The temperature can 
be determined by solving Eq.~(\ref{eq:equip3N}) for $T$.
Figure~\ref{fig:Toft} shows an example
of a MD simulation using positions initially equilibrated at $T=0.5$, and then run at $T=0.2$. The Python program, {\tt KALJ\_nvt.py}, for this section is available at Ref.~\onlinecite{supplement}.

\section{Implementation of MD simulations with LAMMPS}
\label{sec:LAMMPS}

Readers might have noticed that the simulation  
we have discussed was for only $1000$ particles and   
$50$ time steps and 
took painfully long. (The pain level 
depends on the  computer.) To shorten the computation 
time, there exist  
optimization techniques such as    
nearest neighbor lists, as well as coding for multiple 
processors. We now introduce the 
free and open source software, LAMMPS. LAMMPS (Large-scale Atomic/Molecular Massively Parallel Simulator)
allows a very wide range of simulation techniques and 
physical systems.
The  LAMMPS website\cite{LAMMPSwebpage} includes an overview, tutorials, 
well written manual pages,  and links for  downloading  it for a  variety of operating systems.
The goal of this section is to
help  readers get started with LAMMPS and is not a thorough 
introduction to LAMMPS.

\subsection{Introduction to LAMMPS}
\label{sec:LAMMPSintro}

In LAMMPS the user chooses the simulation technique, system,
particle interactions, and parameters, all via an input file.
The main communication between the user and  LAMMPS  
occurs  via the input file.
To run LAMMPS with parallel code, the simulation is started with
commands such as 
\begin{verbatim}
mpirun -np 16 lmp_mpi < inKALJ_nve > outLJnve
\end{verbatim}
where \verb+-np 16+ specifies the number of cores, 
\verb+lmp_mpi+ is the name of the LAMMPS executable 
(which might have a different name depending on the
computer), \verb+outLJnve+ is the output file
(see the following for a description on what information 
is written into this file), and \verb+inKALJ_nve+ 
is the input file.  Becoming familiar with LAMMPS 
mainly requires learning the commands in 
this input file. Further documentation can be found
at Ref.~\onlinecite{LAMMPSwebpage}. A set 
of input file examples is  available at
Ref.~\onlinecite{LAMMPSdownload}. Appendix~\ref{sec:LAMMPSbatchsystem} describes how to run LAMMPS on a shared computer using a batch system.

\subsection{NVE  simulation with LAMMPS}
\label{sec:LAMMPNVE}

To run at fixed NVE 
the input file, \verb+inKALJ_nve+, contains
\begin{widetext}
\begin{verbatim}
#KALJ NVE, read data
atom_style atomic  
boundary p p p #periodic boundary cond. in each direction
read_data initconf_T05eq.data #read data file (incl.mass)
pair_style  lj/cut 2.5  # Define interaction potential
pair_coeff  1 1  1.0 1.0  2.5  #type type eps sigma rcut
pair_coeff  1 2  1.5 0.80 2.0  #typeA typeB epsAB sigmaAB rcutAB=2.5*0.8=2.0
pair_coeff  2 2  0.5 0.88 2.2  #typeB typeB epsBB sigmaBB rcutBB=2.5*0.88=2.2
timestep 0.005 #Delta t 
neighbor          0.3 bin
neigh_modify      every 1 delay 0 check yes 
dump mydump all custom 50 confdump.*.data id type x y z vx vy vz
dump_modify mydump sort id
# set numerical integrator
fix nve1 all nve # NVE; default is velocity verlet
run 100\end{verbatim}
\end{widetext}

Comments start with \verb+#+. The statement \verb+atom_style atomic+ specifies 
the type of particle, and  \verb+boundary p p p+ implements periodic 
boundary conditions.
Not included in this sample input file are the  two possible 
commands, 
\begin{verbatim}
units    lj 
dimension 3
\end{verbatim}
because they are the default settings.

Initial positions and velocities are read  from the file
\verb+initconf_T05eq.data+.
For the LAMMPS \verb+read_data+ command, the specified file 
(here \verb+initconf_T05eq.data+) contains 
\begin{verbatim}
#bin. KALJ data file T=0.5 

1000 atoms
2 atom types

0 9.4 xlo xhi
0 9.4 ylo yhi
0 9.4 zlo zhi

Masses

1 1.0
2 1.0

Atoms

1 1 2.24399 2.3078 9.07631
2 1 8.54631 2.43192 8.67359
...
1000 2 6.99911 8.89427 6.16712

Velocities

1 0.195617 1.29979 -1.17318
2 -0.905996 0.0649236 0.246998
...
1000 -0.661298 -1.71996 2.00882
\end{verbatim}
The first few lines specify the type of system, $N=1000$ atoms 
with $A$ and $B$ particles, the box length, $L=9.4$, and the masses $m_A=m_B=1$.
The $1000$ lines following \verb+Atoms+ specify 
the particle index, $i=1,2,\ldots, 1000$ 
in the first column,  
the particle type in the second column; that is, \verb+1+ for particles
$1, 2, \ldots, 800$ ($A$-particles) and \verb+2+ for
particles $801, \ldots, 1000$ ($B$-particles). 
Columns three, four, and five are $x_i$, $y_i$, $z_i$ respectively.
The lines following \verb+Velocities+ contain $i$, $v_{x,i}$, $v_{y,i}$, 
and $v_{z,i}$.

In the input file \verb+inKALJ_nve+ the particle interactions 
are defined by the  commands \verb+pair_style+ and
\verb+pair_coeff+. Note that \verb+lj/cut+ corresponds 
to the forces of  Eq.~(\ref{eq:KAFix}).
However, the potential
energy excludes the term $V_{\alpha \beta}(r^{\rm cut}_{\alpha \beta})$ 
of Eq.~(\ref{eq:VshiftedKALJ}). In LAMMPS there is also the 
option of the truncated and force shifted Lennard-Jones interactions
\verb+lj/smooth/linear+. We chose  \verb+lj/cut+ to 
allow for the direct comparison of the Python  and 
LAMMPS simulations.
In the file \verb+inKALJ_nve+ the line \verb+timestep 0.005+
sets $\Delta t = 0.005$. The commands \verb+neighbor+ and \verb+neigh_modify+
are parameters for the neighbor list.
The LAMMPS commands \verb+dump+ and \verb+dump_modify+ 
periodically save  snapshots
of all atoms. In our example, 
every 50 time steps (starting with  $t=0$) a file is written with
file name \verb+confdump.0.data+, \verb+confdump.50.data+, \verb+confdump.100.data+,
and the content of each written file has columns $i$, particle type (1 or 2), 
$x_i$, $y_i$, $z_i$, $v_{x,i}$, $v_{y,i}$, and $v_{z,i}$.
In the \verb+dump+ command \verb+mydump+ is the LAMMPS-ID for this 
\verb+dump+ command. It can be replaced with any name the reader   
chooses. The ID allows further specifications for this dump as
used in the command \verb+dump_modify mydump sort id+, which
ensures that the lines in the  dump files are sorted by 
particle index $i$.

The integration technique is set by the  command 
\verb+fix nve1 all nve+; \verb+nve1+ is an ID for this 
\verb+fix+ command, \verb+all+ means that this integration step is 
applied to all particles, and \verb+nve+ specifies the NVE  
time step which is the velocity Verlet integration step by default.
The command \verb+run100+ means that
the  simulation is  run for $100$ time steps under these  
specified conditions.

The input file \verb+inKALJ_nve+  assumes that the initial positions 
and velocities are available. For a small system such as $N=1000$
the initial positions may be generated  by doing a  simulation with
Python. However, for simulations with significantly more 
particles, the initial positions and velocities may not be available.
If we instead initialize with
uniformly randomly distributed positions and
velocities from the Maxwell-Boltzmann distribution, we replace
in \verb+inKALJ_nve+ 
the \verb+read_data+ command with the following LAMMPS commands
\begin{verbatim}
region my_region block 0 9.4 0 9.4 0 9.4
create_box 2 my_region
create_atoms 1 random 800 229609 my_region
create_atoms 2 random 200 691203 my_region
mass 1 1
mass 2 1
velocity all create 0.5 92561 dist gaussian
\end{verbatim}
The first two commands create the simulation box for 
two types of atoms, the \verb+create_atoms+ commands
initialize the atom positions randomly drawn from a 
uniform distribution and random number generator 
seeds \verb+229609+ and \verb+691203+ (any positive integers),
and the last command initializes the velocities of all particles
with the Maxwell-Boltzmann distribution for temperature $T=0.5$ and
random number seed \verb+92561+.

As noted in Sec.~\ref{sec:initialization}, 
 random positions can lead to very large forces. 
These can be avoided by adding in \verb+inKALJ_nve+ 
before the \verb+fix+ command the line
\begin{verbatim}
minimize 1.0e-4 1.0e-6 1000 1000
\end{verbatim}

The files (\verb+inKALJ_nve+,
\verb+inKALJ_nve_rndposvel+, \verb+initconf_T05eq.data+,
and \verb+runKALJ_slurm.sh+) for this section and the previous two sections 
are available.\cite{supplement}

\subsection{NVT  simulations}
\label{sec:LAMMPSNVT}

The default NVT  simulation in LAMMPS 
uses the Nos\'{e}-Hoover algorithm (see Sec.~\ref{sec:NVT_NoseHoover}, 
and Appendices~\ref{sec:derivationNoseHoover} and \ref{sec:FoxAndersonNoseHoover}).
\cite{LAMMPSfixnvtp,shinoda2004,martynaKleinTuckermanJCP1992,martynaMolPhys1996,martynaJCP1994,Frenkel2002}
To implement this temperature bath in LAMMPS, we replace the command 
\verb+fix nve1 all nve+ in
the input file  with
\begin{verbatim}
fix nose all nvt temp 0.2 0.2 $(100.0*dt)
\end{verbatim}
As described in  
Ref.~\onlinecite{LAMMPSfixnvtp},
\verb+nose+ is the ID chosen by the user for this \verb+fix+ command,  \verb+all+ indicates that this \verb+fix+
is applied to all atoms, \verb+nvt temp 0.2 0.2+
sets the constant temperature to $T=0.2$, and the last 
parameter sets the damping parameter as recommended 
in Ref.~\onlinecite{LAMMPSfixnvtp} to $100 \Delta t$.

Another way of  achieving a temperature bath is
to implement the statistical temperature bath 
as described in Sec.~\ref{sec:NVT_Boltzmann}.
We use an implementation similar to that used 
in Python in Sec.~\ref{sec:pythonNVTMD}.
To compute random velocities periodically in time,
we replace the command \verb+run 100+. To  
compute new velocities every $10$ time steps 
at temperature $T=0.2$ the replacement line is 
\begin{widetext}
\begin{verbatim}
run 100 pre no every 10 "velocity all create 0.2 ${rnd} dist gaussian"
\end{verbatim}
\end{widetext}
where \verb+${rnd}+ is a user-defined LAMMPS variable  corresponding to a random reproducible integer; 
\verb+rnd+ needs to be defined before the modified \verb+run+ 
command by
\begin{verbatim}
variable rnd equal floor(random(1,100000,3259))
\end{verbatim}
where we used the LAMMPS function 
\verb+random+ (see Ref.~\onlinecite{LAMMPSvariable}).
To test the program,  readers may plot 
the measured temperature as a function of time, $T_{\rm meas}(t)$ 
(similar to Fig.~\ref{fig:Toft}) and $E_{\rm tot}/N$ 
as a function of time (similar to Fig.~\ref{fig:Etotoft}).
Such time dependent functions can 
be computed and saved
with \verb+thermo_style+,
\begin{verbatim}
thermo_style custom step temp pe ke etotal 
thermo 2 #print every 2 time steps
\end{verbatim}
which saves data every $2$ time steps in the output file, e.g., \verb+outLJnvt+, the
 five variables: (number of time steps), 
$T_{\rm meas}$, $V/N$, $E_{\rm kin}/N$,
and $E_{\rm tot}/N$. Note that the
LAMMPS interaction \verb+lj/cut+ potential
energy excludes the term $V_{\alpha \beta}(r^{\rm cut}_{\alpha \beta})$ 
of Eq.~(\ref{eq:VshiftedKALJ}).

Because the output file includes 
the output from the \verb+thermo+ command plus
several lines with other information, it is 
convenient to filter out the time dependent 
information. This can be done in Unix/Linux. For example, 
to obtain $T_{\rm meas}$ as a function of  time steps, use the Unix command
\begin{widetext}
\begin{verbatim}
gawk 'NF ==5 && ! /[a-z,A-Z]/ {print $1,$2}' outLJnvt
\end{verbatim}
\end{widetext}
The resultant output can be redirected into 
a file or directly piped into a plotting tool, e.g.,
by adding to  \verb+gawk+  at the end
{\tt 
| xmgrace -pipe}.
To obtain $E_{\rm tot}/N$
as a function of the number of time   steps, we 
replace the \verb+gawk+ command \verb+$2+ by \verb+$5+.

The LAMMPS input files, \verb+inKALJ_nvt_stat+ and \verb+inKALJ_nvt_Nose+, are available at Ref.~\onlinecite{supplement}.

\section{Simulation Run Sequence}
\label{sec:SimulRunSequence}

Readers  can now run 
molecular dynamics simulations with Python or 
 LAMMPS.
To illustrate what a simulation sequence entails, we 
give a few examples of simulations for the Kob-Andersen model.

The first set of papers on the Kob-Andersen model were on   the 
equilibrium properties of supercooled liquids.\cite{kobAndersenPRL1994,kobAndersenPRE51_1995,kobAndersenPRE52_1995}
As described in Ref.~\onlinecite{kobAndersenPRE51_1995}, 
the system was first  equilibrated at  $T=5.0$ and then simulated 
at successively lower temperatures 
$T=4.0$, $3.0$, $2.0$, $\ldots$, $0.475$, $0.466$. 
For each successive temperature, a configuration was taken from 
the previously equilibrated temperature run, the temperature
bath (stochastic in this study) was applied for $t_{\rm equi}$
time units, followed by an NVE simulation run also for
$t_{\rm equi}$ time units, and then followed by an NVE 
production run during which the dynamics and structure 
of the system were determined. This  
sequence of reaching successively lower temperatures was 
applied to eight independent initial configurations. 

Another example for a simulation sequence is to 
apply a constant cooling rate as was done in 
Ref.~\onlinecite{KVLKobBinder_1996} with an NPT algorithm.

References~\onlinecite{kobBarratPRL1997,kobBarratPhysicaA1999,kobetalJPhysCondMat2000}
studied the Kob-Andersen model out of equilibrium by first equilibrating the system at a high temperature
$T_{\rm i}$ and then quenching instantly to a lower temperature
$T_{\rm f}$. That is, a well equilibrated configuration 
from the simulation  at $T_{\rm i}$ was taken to be the initial 
configuration for an NVT simulation  at $T_{\rm f}$.
During the run at $T_{\rm f}$ the structure and dynamics 
of the system depend on the waiting time, which is the time elapsed since the temperature quench.
\cite{kobBarratPRL1997,kobBarratPhysicaA1999,kobetalJPhysCondMat2000}

\section{Analysis}
\label{sec:analysis}

In this section we   discuss the
analysis of molecular dynamics simulations.
To give  readers a taste of the wide variety 
of analysis tools, we focus
 on two commonly studied quantities:
the radial distribution function and 
the mean square displacement.

\subsection{Radial distribution function}
\label{sec:gofr}

The radial distribution function, $g(r)$, is an example of
a structural quantity and is a measure of the density of particles $j$ at a distance $r$ from a particle $i$, where 
$r = r_{ij}=\left | {\mathbf r}_i - {\mathbf r}_j \right |$ and 
radial symmetry is assumed. 
For a binary system $g_{AA}(r)$, $g_{BB}(r)$,
and $g_{AB}(r)$ are defined as
\begin{equation}
g_{\alpha \alpha } (r) = \frac{V}{N_\alpha \left (N_\alpha -1 \right )} 
\left \langle \sum \limits_{i=1}^{N_{\alpha}}
 \sum_{\substack{j=1\\ j\ne i}}^{N_{\alpha}}
 \delta \left( r - \left | {\mathbf r}_i - {\mathbf r}_j
 \right | \right)
 \right \rangle ,
\label{eq:gAA}
\end{equation}
where $\alpha \in \{A,B\}$, and (see Refs.~\onlinecite{kobAndersenPRE51_1995,allen90})
\begin{equation}
g_{AB} (r) = \frac{V}{N_{A} N_{B}} 
 \left \langle \sum \limits_{i=1}^{N_{A}}
 \sum \limits_{j=1}^{N_{B}}
 \delta \left( r - \left | {\mathbf r}_i - {\mathbf r}_j
 \right | \right)
 \right \rangle . 
\label{eq:gAB}
\end{equation}
Equations~(\ref{eq:gAA}) and (\ref{eq:gAB}) include sums over 
particle pairs $(i,j)$ of the types specified.
The Dirac delta function $\delta(x)$ is the number 
density for a point particle at $x=0$. The number density of 
$(i,j)$ pairs with distance 
$r=r_{ij}$ is normalized by the global density.
Therefore $g(r)$ characterizes the distribution of particle distances.
The average $\langle \ldots \rangle$  in Eqs.~(\ref{eq:gAA}) and (\ref{eq:gAB}) 
can be taken either by averaging over independent 
simulation runs and/or via a time average by averaging 
measurements at different times $t$.
For the measurement of $g(r)$ in equilibrium, the system needs 
to be first equilibrated, and therefore $t > t_{\rm equil}$ for 
all measurements.
For  more advanced readers the generalization of 
the radial distribution function is the van Hove 
correlation function $G(r,t)$.\cite{HansenVerlet1969,kobAndersenPRE51_1995}

\subsubsection{Radial distribution function with Python}
\label{sec:gofrPython}

\begin{figure}[t]
\begin{center}
\includegraphics[width=\columnwidth]{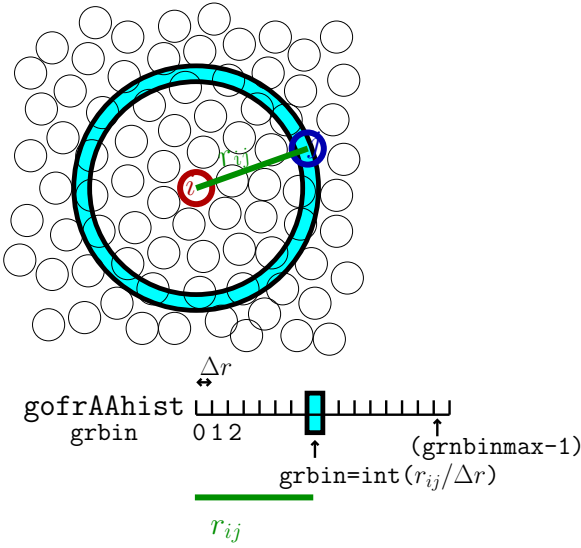}
\vspace{-0.2in}
\caption{Sketch of the determination of the radial 
distribution function. To compute $g_{AA}(r)$ 
a histogram of pair distances $r_{ij}$ is stored in 
the array {\tt gofrAAhist}. The width of each bin is   $\Delta r$.
There are {\tt grnbinmax} bins.}
\label{fig:gofrhist}
\end{center}
\end{figure}

To determine a histogram of the $r_{ij}$ distances, we use 
an array as illustrated for $g_{AA}$ in 
Fig.~\ref{fig:gofrhist}. Before  determining  the histogram we 
set to zero the arrays {\tt gofrAAhist}, {\tt gofrBBhist},
and {\tt gofrABhist}.
For each measurement, we loop over 
all unique particle combinations ($j > i$), determine the 
minimum image distance (see Sec.~\ref{sec:pythonAccelerations}), 
and add to the counter of the 
corresponding bin.\cite{allen90}
\begin{verbatim}
for i in range(0,N-1):
  xi=x[i]
  ...
  for j in range(i+1,N):
    xij=xi-x[j]
    ...
    #minimum image convention
    if xij > Ldiv2: xij -= L
    ...
    rijto2 = xij*xij + yij*yij + zij*zij
    rij=sp.sqrt(rijto2)
    grbin=int(rij/grdelta)
    if(grbin < grnbinmax):
      if(i < Na):
        if (j < Na): #AA
          histgofrAA[grbin] += 2
        else: #AB
          histgofrAB[grbin] += 1
      else: #BB
        histgofrBB[grbin] += 2
\end{verbatim}
Here ${\tt grdelta}=\Delta r$ is the bin size (see   Fig.~\ref{fig:gofrhist}).
If the  average is a time average,
taken via measurements
(assume with user defined function {\tt histmeas})
after {\tt tequil} time steps every {\tt nstepgofr} time steps, 
we set the arrays {\tt histgofrAA} etc.\ to zero before 
the time loop, and  add the conditional statement 
\begin{verbatim}
if (tstep > tequil) and ((tstep % nstepgofr)==0):
   histmeas(x,y,z)
\end{verbatim}
within the time loop and after the time step,
that is in the flow chart of Fig.~\ref{fig:MDflowchartII}
after the (blue) 
``Time Step'' 
box and within 
the ``Loop Over Time Steps,'' so before the (green) time loop
repeats. 
We can then save the resulting radial distribution functions
into a file of name {\tt gofrAABBAB.data}
by adding to the program after the time loop 
\begin{widetext}
\begin{verbatim}
fileoutgofr= open("gofrAABBAB.data",mode='w')
for grbin in range(grnbinmax):
  rinner = grbin*grdelta
  router = rinner + grdelta
  shellvol = (4.0*sp.pi/3.0)*(router**3 - rinner**3)
  gofrAA = (L**3/(Na*(Na-1)))*histgofrAA[grbin]/(shellvol*nmeas)
  gofrBB = (L**3/(Nb*(Nb-1)))*histgofrBB[grbin]/(shellvol*nmeas)
  gofrAB = (L**3/(Na*Nb))*histgofrAB[grbin]/(shellvol*nmeas)
  rmid = rinner + 0.5*grdelta
  print(rmid,gofrAA,gofrBB,gofrAB,file=fileoutgofr)
\end{verbatim}
\end{widetext}
We have assumed that {\tt nmeas} measurements of the histogram were taken, 
and the variables are as shown in Fig.~\ref{fig:gofrhist} with
${\tt grdelta}=\Delta r$. For the normalization we determine 
the shell volume/area, which is sketched in Fig.~\ref{fig:gofrhist}
as the cyan (gray) shaded area enclosed by the two large circles 
drawn with thick lines. In three dimensions 
the shell volume is 
$(4 \pi/3)\left (r_{\rm outer}^3 - r_{\rm inner}^3 \right )$.

\begin{figure}[t]
\begin{center}
\includegraphics[width=\columnwidth]{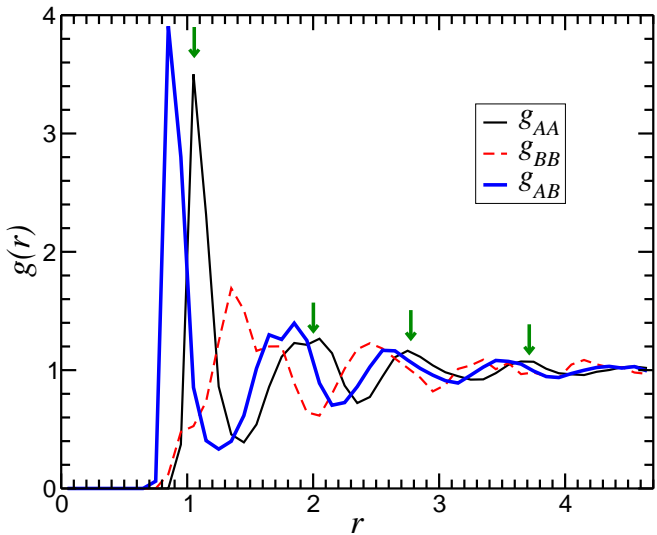}
\caption{Radial pair distribution functions for the Kob-Andersen model
with $N=1000$, $L=9.4$ at $T=0.5$. 
The green vertical arrows indicate peak positions of $g_{AA}(r)$. 
} 
\label{fig:gofrAABBAB}
\end{center}
\end{figure}

The resulting radial distribution functions are 
shown in Fig.~\ref{fig:gofrAABBAB} for which we
ran the NVT  simulation  
with $N=1000$  at $T=0.5$, starting 
with a well equilibrated configuration at $T=0.5$, running the 
simulation for $200$ time steps and measuring the histogram 
every {\tt nstepgofr=25} time steps with $\Delta r=0.1$.
To measure distances up to $L/2$, we set 
\verb+grnbinmax = int(Ldiv2/grdelta)+.
The Python program \verb+KALJ_nvt_gofr.py+ for this section is in Ref.~\onlinecite{supplement}.

\subsubsection{Interpretation of radial distribution function}
\label{sec:gofrInterpretation}

\begin{figure}[t]
\begin{center}
\includegraphics[width= \columnwidth]{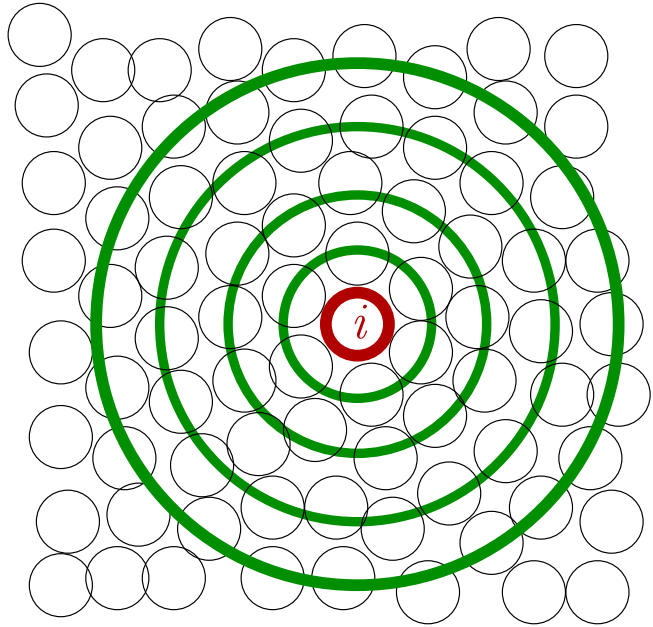}
\caption{Sketch of particle arrangements to illustrate the interpretation 
of the radial distribution function. The first, second, third, and 
forth neighbor shells are indicated with dark green circles (thick 
lines). The radii of these shells correspond to the peak positions of the radial
distribution function as indicated with dark green vertical arrows 
in Fig.~\ref{fig:gofrAABBAB} for the example of $g_{AA}(r)$. 
} 
\label{fig:gofrinterpret}
\end{center}
\end{figure}

Because the repulsive interaction
$V(r_{ij} \to 0) \to \infty$ prevents the complete overlap of 
two particles,  $g(r)=0$. The first peak of $g(r)$ corresponds 
to the most likely radius of the first shell of neighboring 
particles surrounding particle $i$. The second peak of $g(r)$ 
corresponds to the second nearest neighbor shell, etc. 
(see Fig.~\ref{fig:gofrinterpret}). With increasing $r$ the 
peaks become less and less pronounced, because the 
system has, 
contrary to a crystal, no long range order.
The peak positions of $g_{AA}(r)$, $g_{BB}(r)$, and 
$g_{AB}(r)$ in Fig.~\ref{fig:gofrAABBAB} 
are   consistent with the
results of Kob and Andersen (Fig.~9 of Ref.~\onlinecite{kobAndersenPRE51_1995}).
For their more quantitative study, they used longer simulation runs,
 several independent simulation runs, and a
smaller (and probably more than one) value of $\Delta r$.

\subsubsection{Radial distribution function with LAMMPS}
\label{sec:gofrLAMMPS}

We can determine the radial distribution functions with LAMMPS 
by adding to the input file before the \verb+run+ command 
the lines
\begin{widetext}
\begin{verbatim}
compute rdfAABBAB all rdf 25 1 1 2 2 1 2
fix myrdf all ave/time 25 8 200 c_rdfAABBAB[*] file gofrAABBAB.data mode vector
\end{verbatim}
\end{widetext}
The \verb+compute+ command defines the measurements, which
are done during the  run:
\begin{enumerate}
\item {\tt rdfAABBAB} is the user defined ID
for this {\tt compute} command. This ID is used in the {\tt fix} command 
with \verb+c_rdfAABBAB+, which means the ID is like a variable name.

\item {\tt all}   applies the command to all atoms.

\item {\tt rdf} computes the radial distribution function.

\item ${\tt 25}=N_{\rm bin}$ specifies 
$\Delta r$ to be $r^{\rm cut}/25$. The following numbers specify the 
particle type combinations, that is, 
 $g_{AA}$, $g_{BB}$, and $g_{AB}$.

\item  \verb+fix ave/time+  defines the time averaging. 
As described  in Ref.~\onlinecite{LAMMPSfixavetime} 
the three numbers in our example specify that the histogram is measured every
$\mbox{\tt Nevery}=25$ time steps,  ${\tt Nrepeat}=8$ measurements are 
averaged (in Sec.~\ref{sec:gofrPython} \verb+nmeas+), and 
${\tt Nfreq}=200$ is the interval of time steps at which the 
time average is printed. That is, if the simulation run is
$n_{\rm MD}=600$, then the averages of $g(r)$  are printed out 
three times, the first by averaging measurements taken at 
time step $200, 175, 150, \ldots, 25$, and the last one at 
time steps $600, 575, 550, \ldots, 425$. 
Constraints on the choice of \verb+Nevery+, \verb+Nrepeat+,
and \verb+Nfreq+ are given in Ref.~\onlinecite{LAMMPSfixavetime}.
In addition, compatible times need to be chosen,
if the LAMMPS command \verb+run every+ is used, which we 
used for the statistical temperature bath in Sec.~\ref{sec:LAMMPSNVT}.

\item \verb+[*]+ takes time averages for each of the 
variables of the \verb+compute rdf+ command

\item \verb+file gofrAABBAB.data+ specifies  that the results are saved in
the file   \verb+gofrAABBAB.data+.

\item \verb+mode vector+ is 
necessary, because $g_{AA}(r)$ etc. are vectors instead of scalars,
with indices for the $N_{\rm bin}$ bins 
$r=[0,\Delta r), [\Delta r, 2 \Delta r), \ldots$.
The entries in the file \verb+gofrAABBAB.data+ are for each average 
time (here $200$, $400$, $600$) starting with one line
specifying the print time in time steps, $200$ etc., and \verb+Nevery+,
followed by $N_{\rm bin}$ lines, each with columns 
for the bin number, $r$, $g_{AA}$, $c_{AA}$, 
$g_{BB}$, $c_{BB}$, $g_{AB}$, $c_{AB}$,
where $c_{AA}$ etc. are coordination numbers.
\end{enumerate}
The LAMMPS input file \verb+inKALJ_T05_gofr+    is in Ref.~\onlinecite{supplement}.

\subsection{Mean square displacement}
\label{sec:msd}

We next study how the system evolves as a function of time.
The mean square displacement\cite{allen90,kobAndersenPRE51_1995}
captures how far each particle moves during a time interval $t$:
\begin{equation}
\mbox{msd} = \left \langle r^2(t) \right \rangle 
 = \left \langle \left |
 \mathbf{r}(t) - \mathbf{r}(0)
 \right |^2 \right \rangle
\label{eq:msd}
\end{equation}
where $\langle \ldots \rangle$ corresponds to an average 
over particles and may also include 
an average over independent simulation runs.
In the following discussion on the implementation
of the  mean square displacement with Python and  mean square displacement with LAMMPS, 
we  average only over particles of one type
\begin{equation}
\left \langle r^2_{\alpha}(t) \right \rangle 
 = \frac{1}{N_{\alpha}} \sum \limits_{i=1}^{N_{\alpha}} 
 \left | \mathbf{r}_i(t) - \mathbf{r}_i(0) \right |^2,
\label{eq:msdwithi}
\end{equation}
where $\alpha \in \{A,B\}$ is the particle type.

A generalization of Eq.~(\ref{eq:msd}) is 
\begin{equation}
\left \langle r^2(t_{\rm w},t_{\rm w}+t) \right \rangle 
 = \left \langle \left |
 \mathbf{r}(t_{\rm w}+t) - \mathbf{r}(t_{\rm w})
 \right |^2 \right \rangle
\label{eq:msdtw}
\end{equation}
If the system is in equilibrium, 
$\left \langle r^2(t_{\rm w},t_{\rm w}+t) \right \rangle$
is independent of starting time $t_{\rm w}$ and the average 
$\langle \ldots \rangle$ may include an average over $t_{\rm w}$.

\subsubsection{Mean square displacement with Python}
\label{sec:msdPython}

It is suggested that readers do Problem~\ref{prob:msd} before reading the following.  We use arrays to store the positions at $t=0$
after the initialization of the positions with
${\tt x0 = np.copy(x)}$, \ldots. We   cannot  use  
periodic boundary conditions to determine the mean square displacement  and instead use
unwrapped coordinates and define the additional 
arrays {\tt xu}, {\tt yu}, and {\tt zu} which are 
initially also copied from {\tt x}, etc. These 
arrays are updated in the 
time step loop as  in Sec.~\ref{sec:pythonNVEMD}
\begin{verbatim}
  xu += vx*Deltat + 0.5*ax*Deltatto2
  yu += vy*Deltat + 0.5*ay*Deltatto2
  zu += vz*Deltat + 0.5*az*Deltatto2
\end{verbatim}
Periodic boundary conditions are not applied to 
{\tt xu}, {\tt yu}, and {\tt zu}.

To save the results into the file   {\tt msd.data}, 
we add before the time loop the statement
$
\mbox{\tt fileoutmsd = open("msd.data",mode='w')}$.
Measurements of the mean square displacements are 
done within the time loop and after 
the
time step.
\begin{verbatim}
  msdA = 0.0
  for i in range(Na):
    dx = xu[i]-x0[i]
    dy = yu[i]-y0[i]
    dz = zu[i]-z0[i]
    msdA += dx*dx+dy*dy+dz*dz
  msdA /= float(Na)
  msdB = 0.0
  for i in range(Na,N):
    dx = xu[i]-x0[i]
    dy = yu[i]-y0[i]
    dz = zu[i]-z0[i]
    msdB += dx*dx+dy*dy+dz*dz
  msdB /= float(Nb)
  print(tstep*Deltat,msdA,msdB,file=fileoutmsd)
\end{verbatim}

\begin{figure}[t]
\begin{center}
\includegraphics[width=\columnwidth]{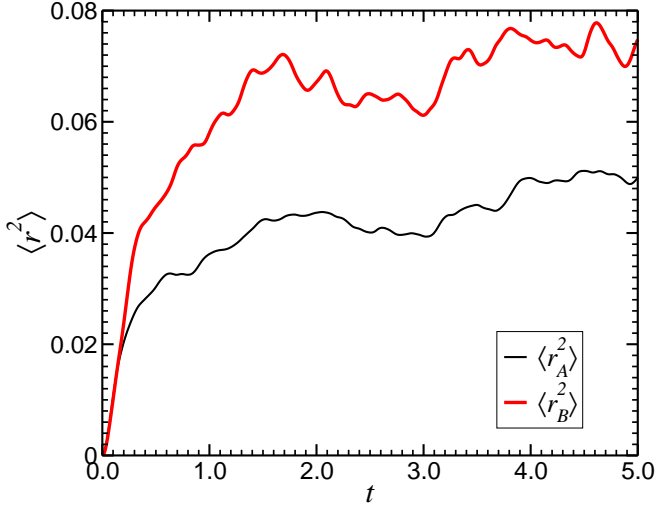}
\vspace{-0.1in}
\caption{The mean square displacement $\langle r^2 \rangle$ 
as function of time $t$ (in LJ units)
for the Kob-Andersen model.
The initial configuration is   equilibrated  
at $T=0.5$. The results are for a NVE simulation 
for $N=1000$  and for $1000$ time steps (with
a Python program, recording $\langle r^2 \rangle$ every 
time step. 
After a steep increase for very small times,
$\langle r^2 \rangle$ reaches a plateau. 
For the interpretation of this figure see Sec.~\ref{sec:msdInterpret}.
} 
\label{fig:msdlin}
\end{center}
\end{figure}

Figure~\ref{fig:msdlin} shows the resultant
mean square displacement as a function of time.
After a steep increase for very small times,
$\langle r^2 \rangle$ reaches a plateau. The 
plateau value is larger for the smaller $B$ particles.
For significantly longer times $\langle r^2 \rangle$ 
increases again. To quantify $\langle r^2(t) \rangle$ 
we  need to record  every 
time step for short times and  longer and longer 
time intervals for larger times so that
 the data points on the horizontal axis are evenly spaced on a log-log plot of $\langle r^2(t) \rangle$
as shown in Fig.~\ref{fig:msdloglong} for a simulation using LAMMPS, which is needed for such larger times.   
This is achieved by  saving data 
at times 
$t_k = t_0 * A^k$. In terms of time steps
\begin{equation}
 \frac{t_k}{\Delta t} = \frac{t_0}{\Delta t} * A^k.
 \label{eq:logtimessteps}
\end{equation}
For $k_{\max}$ print times, we solve 
Eq.~(\ref{eq:logtimessteps}) for $A$ for the case 
of $k=k_{\max}$, when $\left(t_{k_{\rm max}}/\Delta t \right)=n_{\rm MD}$
is the total number of time steps.
\begin{equation}
 A = \left ( \frac{n_{\rm MD}}{\left(t_0/\Delta t \right)} \right )^{(1/k_{\max})}
 \label{eq:A}
\end{equation}
The parameters needed for the calculation of $\langle r^2(t) \rangle$ can be set in Python for the example of
$n_{\rm MD}=1000$, $\left(t_0/\Delta t \right)=1.0$, and $k_{\max}=60$ with 
the following lines before the time step loop:
\begin{verbatim}
kmsdmax = 60 
t0msd = 1.0 
A=(float(nMD)/t0msd)**(1.0/float(kmsdmax)) 
tmsd = t0msd 
tmsdnextint = int(round(t0msd))
\end{verbatim}
where \verb+t0msd+$=\left(t_0/\Delta t \right)$, 
and \verb+tmsd+$=\left(t_k/\Delta t\right)$.
Within the time step loop we add the 
conditional statements
\begin{verbatim}
  if tmsdnextint == tstep:
    # prepare when next msd-time
    while(tmsdnextint == tstep):
      tmsd = A*tmsd
      tmsdnextint = int(round(tmsd))
    # do measurement
    msdA = 0.0
    for i in range(Na):
      dx = xu[i]-x0[i]
    ...
\end{verbatim}
where \verb+...+ continues as above for the msd linear in time.
The \verb+while+ loop was added, because 
for short times \verb+A*tmsd+ might increase 
by less than the integer 1.

The Python programs, \verb+KALJ_nve_msd_lin.py+
and \verb+KALJ_nve_msd_log.py+, for this section are  available in Ref.~\onlinecite{supplement}.

\subsubsection{Mean square displacement with {\tt LAMMPS}}
\label{sec:msdLAMMPS}

The determination of the mean square displacement
requires a computation during the simulation run.
This computation can be done in
LAMMPS with the \verb+compute+ command. (Another example 
of the \verb+compute+ command is in Sec.~\ref{sec:gofrLAMMPS}
for the computation of $g(r)$.)
In Eq.~(\ref{eq:msdwithi}) the sum is
over only $A$ or $B$ particles. Thus, 
in the LAMMPS input script we need to define 
these groups of atoms, which we then use for the 
following \verb+compute+ commands:
\begin{verbatim}
group A type 1
group B type 2
compute msdA A msd 
compute msdB B msd
\end{verbatim}
If we wish to save the mean square displacement 
every time step, or more generally with linear time averaging,
we can use the \verb+fix ave/time+ command as described 
in Sec~\ref{sec:gofrLAMMPS}. 
To write 
$\langle r_A^2 \rangle$ and $\langle r_B^2 \rangle$ 
for every time step  
into files   \verb+msdA.data+ and \verb+msdB.data+,
respectively, we use the  commands
\begin{widetext}
\begin{verbatim}
fix msdAfix A ave/time 1 1 1 c_msdA[4] file msdA.data
fix msdBfix B ave/time 1 1 1 c_msdB[4] file msdB.data
\end{verbatim}
\end{widetext}
The resulting files can  be used to make a figure. 
However, as will become clear in  
Sec.~\ref{sec:msdInterpret},
if long simulation runs of the order of $10^7$ time steps 
are desired,  saving in logarithmic time
becomes necessary (see Sec.~\ref{sec:msdPython}).
Logarithmic printing can be achieved by using
the   function\cite{LAMMPSvariable,logfreq3} \verb+logfreq3+
to define the print times \verb+tmsd+ with the \verb+variable+ 
command and then by using \verb+thermo_style+ and
\verb+thermo+ (see Sec.~\ref{sec:LAMMPSNVT}) 
to print the mean square 
displacements into the output file together with other 
scalar quantities which depend on time.
The previous \verb+fix ave/time+ commands are replaced by
\begin{widetext}
\begin{verbatim}
variable tmsd equal logfreq3(1,200,10000000)
variable tLJ equal step*dt
thermo_style custom v_tLJ c_msdA[4] c_msdB[4] pe etotal 
thermo v_tmsd 
\end{verbatim}
We also defined the variable \verb+tLJ+ for the 
printing of $t$ in LJ units (instead of time steps).\cite{logpython}

Another way to obtain information 
logarithmic in time is to print all unwrapped particle 
positions during the LAMMPS simulation, 
\begin{verbatim}
variable tmsd equal logfreq3(1,200,10000000)
dump msddump all custom 5000 posudump.*.data id xu yu zu
dump_modify msddump sort id every v_tmsd
\end{verbatim}
and then analyze the resulting \verb+posudump*+ files with 
Python or another programming language.
The LAMMPS input files, \verb+inKALJ_nve_msd_lin+, 
\verb+inKALJ_nve_msd_log+, and \verb+inKALJ_nve_msd_logdumps+, for this section are in Ref.~\onlinecite{supplement}.
\end{widetext}

\subsubsection{Interpretation of mean square displacement}
\label{sec:msdInterpret}

\begin{figure}[t]
\begin{center}
\includegraphics[width=\columnwidth]{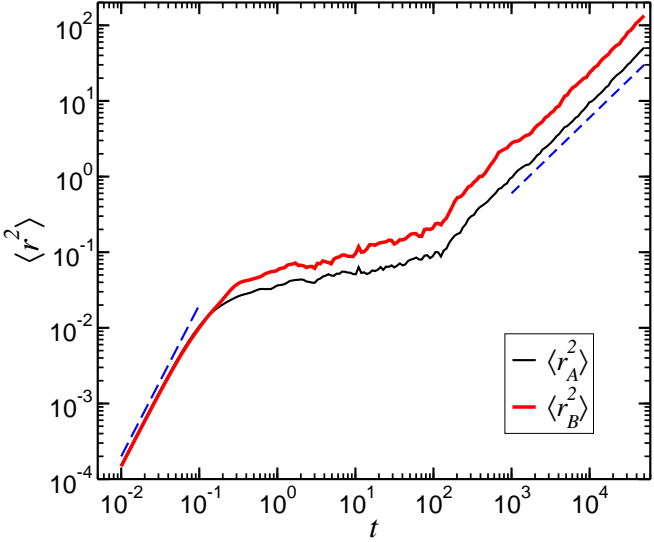}
\vspace{-0.1in}
\caption{The mean square displacement $\langle r^2 \rangle$ 
as a function of the time $t$ (in LJ units) for the Kob-Andersen model.
The initial configuration is  equilibrated  
at $T=0.5$. The results are for the NVE simulation 
with $N=1000$  and for $10^7$ time steps
(with LAMMPS and saving $\langle r^2 \rangle$ 
in logarithmic intervals.)
} 
\label{fig:msdloglong}
\end{center}
\end{figure}

Figure~\ref{fig:msdloglong} shows $\langle r_A^2 \rangle$ and
$\langle r_B^2 \rangle$ obtained with \verb+thermo_style+ 
as described in Sec.~\ref{sec:msdLAMMPS}.

For very short times $t$, we can approximate
$
\mathbf{r}_i(t) = \mathbf{r}_i(0) + \mathbf{v}_i(0)\,t$ and
write Eq.~(\ref{eq:msdwithi})
for small $t$ as
\begin{equation}
\langle r_{\alpha}^2 \rangle  =  \frac{1}{N_{\alpha}} 
 \sum \limits_{i=1}^{N_{\alpha}} \left | \mathbf{v}_i(0) \, t \right |^2 =  C \, t^2 .
\end{equation}
We see that $\ln \left ( \langle r_{\alpha}^2 \rangle \right )
 = \ln C  +  2 \ln t$,
corresponding to a line with slope $2$ as indicated by the 
 dashed line at short times in Fig.~\ref{fig:msdloglong}.

For intermediate times $\langle r_{\alpha}^2 \rangle$
reaches a plateau. This   plateau 
is typical for glass formers
at high enough density at which each particle is trapped 
in a cage formed by its neighboring particles. The smaller $B$ particles 
reach a higher plateau. For long enough times each particle 
escapes its cage of neighbors and therefore 
$\langle r_{\alpha}^2 \rangle$ increases. At very large 
times the dynamics of many successive escape events can be modeled as a 
random walk. For a random walk in $d$ dimensions of step size $a$ and an equal 
probability to step right or left, we have
 after $N_{\rm step}$
steps\cite{gouldTobochnikChristian2007}
\begin{equation}
\langle \left ( \Delta \mathbf r \right )^2 \rangle = d  a^2  N_{\rm step}.
\label{eq:short}
\end{equation}
Equation~\eqref{eq:short} implies that $\langle r_{\alpha}^2 \rangle \propto t$
and therefore a  log-log plot yields a line of slope $1$ as indicated by
a blue dashed line at long times in Fig.~\ref{fig:msdloglong}.
 
\section{Suggested problems}
\label{sec:problems}

\begin{enumerate}
 \item \label{prob:Fix} Determine $F_{i,x} = - {\rm d}V/{\rm d}x_i$.
 
 \item \label{prob:MDflowchart}Sketch the flow chart for the
molecular dynamics simulation in Fig.~\ref{fig:MDintro} in more detail,
specifying the order of the determination of positions,
velocities, accelerations, and the application of periodic boundary conditions.

 \item \label{prob:EkinTrelation} To derive Eq.~(\ref{eq:equip3N}), first 
    determine $\langle \frac{1}{2} m_i v_{i,x}^2 \rangle =
              \!\int \limits_{-\infty}^{\infty} \frac{1}{2} m_i 
                   v_{i,x}^2 P\left(v_{i,x}\right) {\rm d}v_{i,x}$
   (the result is a special case of the equipartition theorem), and then
   obtain Eq.~(\ref{eq:equip3N}).
   
 \item \label{prob:poslattice} Write a program that places the $N$ 
                particles on lattice sites of a simple cubic lattice.
 \item \label{prob:accfct} Outline the implementation of 
        the acceleration function and program it with Python.
        
 \item \label{prob:NVEMD} Use Fig.~\ref{fig:MDflowchartII} to add to 
         your Python program the loop over time steps and 
         update positions and velocities.
 \item \label{prob:NVTstochastic} Use Sec.~\ref{sec:NVT_Boltzmann} 
         to add to your Python program the stochastic temperature bath algorithm.
 \item \label{prob:msd} Add to your NVE Python program the determination
  of the mean square displacement and save the results in a file.
 \item \label{prob:Hderivation} For the $(3N+1)$ generalized coordinates 
      $\mathbf{q}=\left(\left\{\mathbf{r}_i\right\},s \right)$ 
        determine the conjugate momenta $\mathbf{p}_i$ and $p_s$ and 
      then the Hamiltonian.
 \item To simulate a system of  sheared bubbles, Durian\cite{durianPRL1995,durianPRE1997} introduced 
       a model such that bubble $i$ of radius $R_i$ interacts 
       with bubbles $j$ of radius $R_j$ as
 \begin{equation}V_{ij}=\frac{F_0}{2} \left [ 
              1 - \frac{r_{ij}}{\left (R_i+R_j \right )}
                             \right ]^2
  \end{equation}
       for all $j$ with $r_{ij} \le \left (R_i+R_j \right )$.
       Determine the force $\mathbf{F}_{ij}$ on particle $i$ 
       due to particle $j$.
       The solution is Eq.~(1) in Ref.~\onlinecite{durianPRE1997}.
       
 \item Compute the radial distribution functions $g_{\alpha \beta}$
       for (a) temperatures
       $0.1 \le T \le 3.0$ and  (b) densities 
       $0.1 \le N/L^3 \le 2.0$.
       For each parameter set,   first equilibrate before 
         measuring $g_{\alpha \beta}$. Choose $\Delta r \le 0.05$.
       In Python you can include a parameter in the name of the 
       output file. For example, to use the temperature in the name we can write
       \begin{widetext}
       \verb-fileoutgofr= open("gofrAABBAB"+str(temperature)+".data",mode='w')-.
       \end{widetext}
       Interpret your results. 
       Reference~\onlinecite{kobAndersenPRE51_1995} includes $g_{\alpha \beta}$
       for $0.466 \le T \le 5.0$ in Fig.~9 as well as a discussion of its behavior.
       
 \item Compute the mean square displacement given in 
       Eq.~(\ref{eq:msdtw}). Average  over each type of particle separately, that is, compute
  \begin{equation}\left \langle r^2_{\alpha}(t_{\rm w},t_{\rm w}+t) \right \rangle 
       =\frac{1}{N_{\alpha}} \sum \limits_{i=1}^{N_{\alpha}} \left | 
       \mathbf{r}_i(t_{\rm w}+t) - \mathbf{r}_i(t_{\rm w}) \right |^2.
\end{equation}
Use several values of $t_{\rm w}$. For example in a  run with 
       $n_{\rm MD}=1000$, use 
       $t_{\rm w}=0, 10 \Delta t, 100 \Delta t, 500 \Delta t$. 
       First do a NVE simulation as  done in Sec.~\ref{sec:msd}.
       Use as initial configuration the provided 
       file \verb+initposvel+ (for Python) or \verb+initconf_T05eq.data+ 
       (for LAMMPS), which is well equilibrated at $T=0.5$.
       Make a plot of 
       $\left \langle r^2_{\alpha}(t_{\rm w},t_{\rm w}+t) \right \rangle$ 
       as a function of the time difference $t$ for  different values of $t_{\rm w}$. 
       Interpret your results. 
       Then, using the same initial configuration file, 
        do a NVT simulation at $T=0.2$.  
       Make again  a plot of
       $\left \langle r^2_{\alpha}(t_{\rm w},t_{\rm w}+t) \right \rangle$
       as a function of the time difference $t$ for  different values of 
       $t_{\rm w}$. Compare your plots for the NVE run ($T=0.5$)  
       and the NVT run at $T=0.2$ and 
       interpret your results.

 \item Determine the mean square displacement 
       $\left \langle r^2_{\alpha}(t) \right \rangle$ of the KA-LJ system 
       for $N=1000$, $L=9.4$ at temperatures $3.0, 2.0, 1.0, 0.8, 0.6,
       0.55, 0.5,$ and $0.475$ (a subset of the temperatures studied by
       Kob and Andersen.\cite{kobAndersenPRE51_1995} 
       Be sure to equilibrate
       the system sufficiently at each investigated temperature.
       As in Ref.~\onlinecite{kobAndersenPRE51_1995} start at
       $T=3.0$, apply a temperature bath for $n_{{\rm equi},T=3}$ time steps,
       continue with a NVE simulation run of $n_{{\rm equi},T=3}$ time steps,
       use the resulting configuration as initial configuration for the 
       production run at $T=3.0$ and also as initial configuration 
       of the next lower temperature, $T=2.0$. Apply the temperature 
       bath  at $T=2.0$ for $n_{{\rm equi},T=2}$ time steps,
       followed by a NVE simulation run of $n_{{\rm equi},T=2}$ time steps,
       etc. For $\Delta t=0.0025$ we recommend $n_{{\rm equi},T}=10^6$
       time steps for $T \ge 0.8$, 
       $n_{{\rm equi},T}=2 \times 10^6$ for $0.6 \ge T \ge 0.55$, and
       $n_{{\rm equi},T}=5 \times 10^6$ for $0.5 \ge T \ge 0.475$.  
       When  doing  a sequence of NVT and NVE runs, use 
       the LAMMPS command \verb+unfix+ before applying the next \verb+fix+ 
       command. To be able to apply  logarithmic printing 
       of the mean square displacement as in Sec.~\ref{sec:msdLAMMPS}, 
       you may also use the LAMMPS command \verb+reset_timestep 0+.
       To ensure that the neighbor list is updated sufficiently frequently,
       use the LAMMPS command \verb+neighbor 0.2 bin+ (instead of 
       \verb+neighbor 0.3 bin+).
       If the Nos\'{e} temperature bath is used, we  recommend
       for $T \le 0.5$ to scale the velocities after the 
       NVT run such that the total energy per particle of the NVE run 
       is equal to
       the average total energy per particle  $\langle E_{\rm tot}/N\rangle$
       during the NVT run with the (time) average 
       taken near the end of the NVT run.
       Velocity scaling can be achieved with the LAMMPS command 
       \verb+velocity all scale ${scaleEn}+, 
       where \verb+{scaleEn}+ corresponds to the temperature corresponding
       to the time averaged total energy per particle obtained for example 
       with the LAMMPS commands
\begin{widetext}
\begin{verbatim}
variable etot equal ke+pe 
fix aveEn all ave/time 1 500000 2000000  v_etot
variable scaleEn equal (2*(f_aveEn-pe))/3
\end{verbatim}
\end{widetext}
       These commands need to be before the \verb+run+ command of the NVT run.
       Interpret the resulting mean square displacements and 
       compare your results with Fig.~2 of
       Ref.~\onlinecite{kobAndersenPRE51_1995}, keeping in mind,
       that we use the time unit $\sqrt{m_A \sigma_{AA}^2/\epsilon_{AA}}$,
       whereas Kob and Andersen use the time unit 
       $\sqrt{m_A \sigma_{AA}^2/48 \epsilon_{AA}}$.
       For large times $t$ the mean square displacement depends on $t$ as   (see Ref.~\onlinecite{gouldTobochnikChristian2007})
       \begin{equation}
          \left \langle r^2_{\alpha}(t) \right \rangle = 2 d D_{\alpha} t,
          \label{eq:msd6Dt}
       \end{equation}
   where 
 $D_{\alpha}$ is the diffusion constant
       for particles $\alpha \in \{A,B\}$.
       Determine $D_{\alpha}(T)$ by fitting Eq.~(\ref{eq:msd6Dt}) to
       $\left \langle r^2_{\alpha}(t) \right \rangle$.
       Fitting 
       can be done for example with Python        or  gnuplot. For each fit check the goodness of the fit by
       eye by plotting your data and the fitting curve. 
       To ensure that Eq.~(\ref{eq:msd6Dt}) is a good approximation
       adjust the $t$-range used for the fitting accordingly.
       Use the resulting fit parameters to obtain $D_A(T)$ and $D_B(T)$.
       As done in Ref.~\onlinecite{kobAndersenPRE51_1995},
       fit the predictions from mode-coupling theory
       \begin{equation}
            D_{\alpha} = A \left ( T - T_{\rm c} \right)^{\gamma_{\alpha}},
       \end{equation}
       and check your fits with a log-log plot of $D_{\alpha}$ as function 
       of $(T-T_{\rm c})$. Another prediction for $D_{\alpha}(T)$ is the 
       Vogel-Fulcher law
       \begin{equation}
            D_{\alpha} = C \exp \left [-B/( T - T_{\rm VF}\right] \,.
       \end{equation}
       Compare your results with Fig.~3 of 
       Ref.~\onlinecite{kobAndersenPRE51_1995} which shows 
       $6 D_{\alpha}$ (not $D_{\alpha}$ \cite{DKAFig3})
       with diffusion unit $\sigma_{AA} \sqrt{48 \epsilon_{AA}/m_A}$
       (instead of $\sigma_{AA} \sqrt{\epsilon_{AA}/m_A}$ as for your results).
       The results are discussed in 
       Ref.~\onlinecite{kobAndersenPRE51_1995}.
       
 \item Simulate the binary Lennard-Jones system in two  instead of three dimensions. Choose the 
       same density $N/L^2=1.204$ with  $N=1000$ and 
 $L=28.82$. Either start with random positions 
       and velocities from a Maxwell-Boltzmann distribution, or 
       use the input file configurations 
        \verb+initposvel_2d_lammps.data+ (for LAMMPS) or \verb+initposvel_2d_python.data+
       (for Python).
       Both are a result of simulations at $T=0.2$ and are 
       provided in Ref.~\onlinecite{supplement}. Do an NVE 
       or  NVT simulation for $T<0.5$.
       Remember to 
       replace   $3$ 
       by $2$ in Eq.~(\ref{eq:equip3N}) and  adjust the variable \verb+shellvol+ in the Python computation of the radial distribution function.
       For LAMMPS follow the instructions in Ref.~\onlinecite{LAMMPS2d};
        you may set $z_i=v_{i,z}=0.0$ 
       with the LAMMPS command
       \verb+set atom 1000 z 0.0 vz 0.0+.
       Make a scatter plot of the resulting particle positions  
       and compute the radial distribution function. 
       Compare with
       the three-dimensional results. An interpretation 
       of your results is given   in Ref.~\onlinecite{KALJ2dBruening}, which  introduced 
       the two-dimensional Kob-Andersen Lennard-Jones model with
       the particle ratio $N_A:N_B=65:35$ instead of $80:20$.      
\end{enumerate}

\acknowledgments
I thank my former advisors 
W.~Kob and K.~Binder, who introduced me to molecular dynamics 
simulations when I was a student.
I am thankful to J.~Horbach, G.~P.~Shrivastav,
Ch.~Scherer, E.~Irani, B.~Temelso, and T.~Cookmeyer for introducing
me to LAMMPS.
I am also grateful to my former students in my 
research group as well as my computer simulations course, 
in particular, T.~Cookmeyer, L.~J.~Owens, S.~G.~McMahon, 
K.~Lilienthal, J.~M.~Sagal, and M.~Bolish, 
for their questions, which 
were a guidance for this paper. 
I thank   my department for their expertise and 
passion in teaching. Some of our advanced lab materials 
influenced this paper.
I thank the Institute of Theoretical
Physics in G\"ottingen and P.~Sollich for hosting me 
during my sabbatical when 
major parts of this paper were written.

\appendix

\section{Hamiltonian Formalism for Nos\'{e}-Hoover Thermostat}
\label{sec:derivationNoseHoover}

We  motivate Eqs.~(\ref{eq:nosehoover_rddot}) 
and (\ref{eq:nosehoover_xidot})   using the 
Hamiltonian formalism. We follow the derivation given 
in Chapter 6 of Ref.~\onlinecite{Frenkel2002} and present a shortened version here
for simplicity.
For a  complete derivation see Refs.~\onlinecite{Frenkel2002,nosehoover,branka}.

We start with the Lagrangian 
\begin{eqnarray}
 \label{eq:NoseLagrangian}
 {\cal L} & = & \sum \limits_{i=1}^{N} \frac{1}{2} m_i 
 \left ( s \, \dot{\mathbf{r}}_i \right )^2 
 - V \left (\{\mathbf{r}_i\} \right ) \nonumber \\
 & &{} + \frac{1}{2} Q \dot{s}^2 - X k_{\rm B} T \ln s,
\end{eqnarray}
where $X=dN$  
(see Problem~\ref{prob:Hderivation}).
The  momenta are 
$p_{i,x} = \partial {\cal L}/\partial \dot{r}_{i,x}
 = m_i s^2 \dot{r}_{i,x}
$ and similarly for $p_{i,y}$ and $p_{i,z}$. Therefore
\begin{equation}
 \mathbf{p}_i = m_i s^2 \dot{\mathbf{r}}_i 
\label{eq:pi} 
\end{equation}
Similarly,
$p_s = \partial {\cal L}/\partial \dot{s}
 = Q \dot{s}$.
We apply Hamiltonian mechanics\cite{Taylor} using as generalized
coordinates $q_k$ for $k=1, 2, \ldots 3 N+1$, where the first $3N$ values 
of $k$ label
$q_k=r_{i,\mu}$ for particles $i=1,\ldots,N$ and  
$\mu \in \{x,y,z\}$ and $q_{3N+1}=s$.
The corresponding Hamiltonian is
\begin{align}
{\cal H} & =  \sum \limits_{k=1}^{3N+1} \dot{q}_k \, p_k - {\cal L} \\
 & =  \sum \limits_{i=1}^N \frac{\mathbf{p}_i^2}{2 m_i s^2} 
 + V \left (\{\mathbf{r}_i\} \right )
 + \frac{p_s^2}{2 Q} + X k_{\rm B} T \ln s \label{eq:HNoseFirst}
\end{align}
We use Hamilton's equations $\dot{q}_k  =  \partial {\cal H}/\partial p_k$ and $ \dot{p}_k  =  - \partial {\cal H}/\partial q_k$ to 
 obtain the equations of motion
\begin{align}
 \dot{\mathbf{r}}_i & =  \frac{\partial {\cal H}}{\partial \mathbf{p}_i} 
 = \frac{\mathbf{p}_i}{m_i s^2} \label{eq:ridot}\\
 \dot{s} & =  \frac{\partial {\cal H}}{\partial p_s} = \frac{p_s}{Q}
 \label{eq:sdot}\\ 
 \dot{\mathbf{p}}_i & =  - \frac{\partial {\cal H}}{\partial \mathbf{r}_i} 
 = - \nabla_i V = \mathbf{F}_i 
 \label{eq:pidot}\\
 \dot{p}_s & =  - \frac{\partial {\cal H}}{\partial s} 
 = \sum \limits_{i=1}^N \frac{p_i^2}{m_i s^3} 
 - \frac{X k_{\rm B} T}{s}.
 \label{eq:psdot} 
\end{align}
We follow Frenkel and Smit\cite{Frenkel2002} and  switch to ``real variables'' 
$\tilde{\mathbf{r}}_i$, $\tilde{\mathbf{p}}_i$, $\tilde{s}$,
$\tilde{p}_s$, ${\rm d}\tilde{t}$, corresponding to a rescaling of the time:
\begin{align}
 \tilde{\mathbf{r}}_i & =  \mathbf{r}_i \label{eq:rtilde}\\
 \tilde{\mathbf{p}}_i & =  \frac{\mathbf{p}_i}{s} \label{eq:ptilde}\\
 \tilde{s} & =  s \label{eq:stilde} \\
 \tilde{p_s} & =  \frac{p_s}{s} \label{eq:pstilde} \\
 {\rm d}\tilde{t} & =  \frac{{\rm d}t}{s} \label{eq:dttilde}
\end{align}
We also define 
\begin{equation}
 \xi = \frac{{\rm d}s}{{\rm d}t} 
 = \frac{{\rm d} \ln \tilde{s}}{{\rm d} \tilde{t}}
 \label{eq:xidef}
\end{equation}

The equations of motion for the real variables are
\begin{align}
 \frac{{\rm d}\tilde{\mathbf{r}}_i}{{\rm d}\tilde{t}}
 & = \frac{{\rm d}{\mathbf{r}}_i}{\frac{1}{s} {\rm d}{t}}
 = s \frac{\mathbf{p}_i}{m_i s^2}
 = \frac{\tilde{\mathbf{p}}_i}{m_i} \label{eq:rtildedot} \\
 \frac{{\rm d}\tilde{\mathbf{p}}_i}{{\rm d}\tilde{t}}
 & = s \frac{{\rm d}}{{\rm d}t} \left ( \frac{\mathbf{p}_i}{s} \right )
 = \frac{{\rm d}\mathbf{p}_i}{{\rm d}t} 
 - \mathbf{p}_i \frac{\left (\frac{{\rm d}s}{{\rm d}t} \right )}{s} 
 = \mathbf{F}_i - \tilde{\mathbf{p}}_i \xi \label{eq:ptildedot} \\
\frac{{\rm d}\xi}{{\rm d}\tilde{t}} 
 & = \frac{{\rm d}}{{\rm d} \tilde{t}} 
 \left ( \frac{{\rm d}s}{{\rm d}t} \right ) 
 = s \frac{\rm d}{{\rm d}t} \left (\frac{p_s}{Q} \right )  \nonumber \\
 & = \frac{s}{Q} \left ( \sum \limits_{i=1}^N \frac{p_i^2}{m_i s^3}
 - \frac{X k_{\rm B} T}{s} \right )  
= \frac{1}{Q} \left ( \sum \limits_{i=1}^N 
 \frac{\tilde{\mathbf{p}}_i^2}{m_i} - X k_{\rm B} T \right )
 \label{eq:xidot}.
\end{align}
By using Eq.~(\ref{eq:HNoseFirst}) the Hamiltonian in terms of 
real variables can be expressed as
\begin{equation}
 {\cal H} = \sum \limits_{i=1}^N \frac{\tilde{p}_i^2}{2 m_i} 
 + V \left (\{\mathbf{r}_i\} \right )
 + \frac{Q}{2} \xi^2 + X k_{\rm B} T \ln \tilde{s} .
 \label{eq:HNosetilde}
\end{equation}
For the equations of motion (\ref{eq:rtildedot})--(\ref{eq:xidot}) the constant of motion is given by 
Eq.~(\ref{eq:HNosetilde}).
Note that Eqs.~(\ref{eq:ptildedot}) and (\ref{eq:xidot}) are 
the same as Eqs.~(\ref{eq:nosehoover_rddot}) and 
(\ref{eq:nosehoover_xidot}) 
by replacing in Eqs.~(\ref{eq:ptildedot}) and (\ref{eq:xidot}) 
$\tilde{\mathbf{r}}_i, \tilde{s}, \tilde{\mathbf{p}}_i, \tilde{p}_s$ with
$\mathbf{r}_i, s, \mathbf{p}_i, p_s$; that is, we do a 
(confusing) change of notation for the sake of simplicity in 
Sec.~\ref{sec:NVT_NoseHoover}.

\section{Fox-Anderson Integration of the Nos\'{e}-Hoover Equations}
\label{sec:FoxAndersonNoseHoover}

We cannot directly apply the velocity-Verlet algorithm of 
Eqs.~(\ref{eq:velVerletPos}) and (\ref{eq:velVerletVel}) to numerically integrate Eqs.~(\ref{eq:nosehoover_rddot}) and
(\ref{eq:nosehoover_xidot}) because the acceleration
$\mathbf{a}_i(t+\Delta t)$ depends on the velocity 
$\mathbf{v}_i(t+\Delta t)$.
We use instead the more general velocity Verlet integration technique 
of Fox and Andersen\cite{fox} and apply it to the NVT Nos\'{e}-Hoover
equations of motion. As described in Appendix~A of Ref.~\onlinecite{fox} 
this technique is applicable when the form of the equations of motion is 
\begin{align}
\ddot{\mathbf{x}}(t) & = 
 f[\mathbf{x}(t),\dot{\mathbf{x}}(t),\mathbf{y}(t),\dot{\mathbf{y}}(t)]
 \label{eq:xddot} \\
 \ddot{\mathbf{y}}(t) & =  g[\mathbf{x}(t),\dot{\mathbf{x}}(t),\mathbf{y}(t)],
 \label{eq:yddot} 
\end{align}
These equations can be expressed  as [see Ref.~\onlinecite{fox}, Eq.~(A.4)]
\begin{widetext}
\begin{align}
 \mathbf{x}(t+\Delta t) & =  \mathbf{x}(t) 
 + \dot{\mathbf{x}}(t) \, \Delta t 
 + 0.5 f[\mathbf{x}(t),\dot{\mathbf{x}}(t),\mathbf{y}(t),\dot{\mathbf{y}}(t)] \left ( \Delta t \right )^2 
 \label{eq:FoxAndersen_xtplush} \\
 \mathbf{y}(t+\Delta t) & =  \mathbf{y}(t) 
 + \dot{\mathbf{y}}(t) \, \Delta t 
 + 0.5 g[\mathbf{x}(t),\dot{\mathbf{x}}(t),\mathbf{y}(t)] \left ( \Delta t \right )^2 
 \label{eq:FoxAndersen_ytplush} \\
 \dot{\mathbf{y}}^{\rm approx} (t + \Delta t) & =  \dot{\mathbf{y}}(t) 
 + 0.5  \left \{ 
 g[\mathbf{x}(t),\dot{\mathbf{x}}(t),\mathbf{y}(t)] 
 + g[\mathbf{x}(t+\Delta t),\dot{\mathbf{x}}(t),\mathbf{y}(t+\Delta t)] 
 \right \}  \Delta t 
 \label{eq:FoxAndersen_ydotapprox} \\
\dot{\mathbf{x}}(t+\Delta t) & =  \dot{\mathbf{x}}(t) + 0.5  \Big \{
 f[\mathbf{x}(t),\dot{\mathbf{x}}(t),\mathbf{y}(t),\dot{\mathbf{y}}(t)] 
\nonumber \\
& \quad 
+ f[\mathbf{x}(t+\Delta t),\dot{\mathbf{x}}(t+\Delta t),
\mathbf{y}(t+\Delta t),\dot{\mathbf{y}}^{\rm approx}(t+\Delta t)] 
 \Big \}  \Delta t 
\label{eq:FoxAndersen_xdottplush} \\
 \dot{\mathbf{y}}(t+\Delta t) & =  \dot{\mathbf{y}}(t) + 0.5  \left \{
 g[\mathbf{x}(t),\dot{\mathbf{x}}(t),\mathbf{y}(t)] 
 + g[\mathbf{x}(t+\Delta t),\dot{\mathbf{x}}(t+\Delta t),\mathbf{y}(t+\Delta t)] 
 \right \}  \Delta t .
 \label{eq:FoxAndersen_ydottplush} 
\end{align}
\end{widetext}
As Fox and Andersen note, Eq.~(\ref{eq:FoxAndersen_xdottplush}) contains
$\dot{\mathbf{x}}(t+\Delta t)$ on both sides. 
For the case of Nos\'{e}-Hoover equations, Eq.~(\ref{eq:FoxAndersen_xdottplush})
can be solved for $\dot{\mathbf{x}}(t+\Delta t)$. 
We write
\begin{align}
 \mathbf{x} & =  \left \{\mathbf{r}_i \right \}\\
 y & =  \ln s \\
 \dot{y} & =  \xi \\
 \ddot{y} & =  \dot{\xi} \\
 f & =  \left \{ \frac{{\mathbf F}_i}{m_i} - \xi \, \dot{\mathbf r}_i 
 \right \} \\
 g & =  
 \frac{1}{Q} 
 \left ( \sum \limits_{i=1}^{N} m_i \dot{{\mathbf r}}_i^2 - X k_{\rm B} T
 \right ).
\end{align}
and see that Eqs.~(\ref{eq:FoxAndersen_xtplush}--(\ref{eq:FoxAndersen_ydotapprox}) correspond to
\begin{widetext}
\begin{align}
 \mathbf{r}_i(t+\Delta t) & = \mathbf{r}_i(t) + \mathbf{r}_i(t) \, \Delta t 
 + 0.5 \left [ \frac{\mathbf{F}_i(t)}{m_i} - \xi \, \mathbf{r}_i(t)
 \right ] \left ( \Delta t \right )^2 \label{eq:Noseupdate_ri} \\
 \ln s(t+\Delta t) & =  \ln s(t) + \xi(t) \Delta t
 + \frac{1}{2 Q} 
 \left ( \sum \limits_{i=1}^{N} m_i \dot{\bf{r}}_i^2(t) - X k_{\rm B} T
 \right )  \left ( \Delta t \right )^2
 \label{eq:Noseupdate_lns}\\
 \xi^{\rm approx}(t+\Delta t) & =  \xi(t) 
 + \frac{\Delta t}{Q}
 \left [ \sum \limits_{i=1}^{N} m_i \dot{\mathbf{r}}_i^2(t) - X k_{\rm B} T
 \right ] \label{eq:Noseupdate_xiapprox}.
\end{align}
Equation~(\ref{eq:FoxAndersen_xdottplush}) corresponds to
\begin{eqnarray}
 \dot{\mathbf{r}}_i(t + \Delta t) &=& \dot{\mathbf{r}}_i(t) +
\frac{1}{2}  \Big \{
 \left [ \frac{{\mathbf F}_i(t)}{m_i} 
 - \xi(t) \, \dot{\mathbf r}_i(t) \right ] \nonumber \\
 &&{}+ \left ( \frac{{\mathbf F}_i(t+\Delta t)}{m_i} 
 - \xi^{\rm approx}(t+\Delta t)  \dot{\mathbf r}_i(t+\Delta t) 
 \right )
 \Big \} \Delta t,
\end{eqnarray}
which can be solved for $\dot{\mathbf{r}}_i(t + \Delta t)$
\begin{equation}
 \label{eq:ridotsolved}
 \dot{\mathbf{r}}_i(t + \Delta t) = \left \{
 \dot{\mathbf{r}}_i(t) + \frac{\Delta t}{2} 
 \left [ \frac{{\mathbf F}_i(t)}{m_i} - \xi(t)\dot{\mathbf r}_i(t)
 + \frac{\mathbf{F}_i(t+\Delta t)}{m_i} \right ]
 \right \}  \left [ 1 + \frac{\Delta t}{2} \xi^{\rm approx}(t+\Delta t) 
 \right ]^{-1}.
\end{equation}
We use a Taylor series and keep terms up to order $(\Delta t)^2$ and 
obtain Eq.~(A9) of Ref.~\onlinecite{KVLRomanHorbach}:
\begin{eqnarray}
 \label{eq:ridotKVLRoman}
 \dot{\mathbf r}_i(t+\Delta t) &=& \dot{\mathbf r}_i(t) 
 + \frac{\Delta t}{2} 
 \bigg \{ \frac{{\mathbf F}_i(t)+{\mathbf F}_i(t+\Delta t)}{m_i} 
 - \big [ \xi(t) 
 + \xi^{\rm approx}(t+\Delta t) \big ] 
  \dot{\mathbf r}_i(t) 
 \bigg \} \nonumber \\
 &&{}\times \bigg [
 1 - \frac{\Delta t}{2} \xi^{\rm approx}(t+\Delta t) \bigg ].
\end{eqnarray}
\end{widetext}

\section{Batch System}
\label{sec:LAMMPSbatchsystem}

This appendix   is necessary only  if the 
reader uses a supercomputer with a batch system.  Often, supercomputers with \verb+mpirun+ do not 
allow the direct, interactive running of programs. Instead a 
batch system is used to provide computing power to many users 
who run many and/or long (hours--months) simulations. In this case an extra step is needed. That is,
the user writes a batch-script, which contains the 
\verb+mpirun+ command, and submits a  run 
via this script. Some of these script commands are supercomputer
specific.

An example of a slurm batch-script is
\begin{verbatim}
#!/bin/bash
#SBATCH -p short # partition (queue)
#SBATCH -n 16 # number of cores
#SBATCH --job-name="ljLammps" # job name
#SBATCH -o slurm.%N.%j.out # STDOUT 
#SBATCH -e slurm.%N.%j.err # STDERR
module load lammps 
# sometimes also mpi module needs to be loaded
mpirun -np 16 lmp_mpi < inKALJ_nve > outLJnve
\end{verbatim}

This script, with file name \verb+runKALJ_slurm.sh+, 
is  submitted with slurm using the command
\begin{verbatim}
sbatch runKALJ_slurm.sh
\end{verbatim}

We can look at  submitted jobs using \verb+squeue+ and if
necessary, kill a submitted job with \verb+scancel+.


\begin{thebibliography}{99}

\bibitem{LAMMPSwebpage}LAMMPS molecular dynamics simulator, \url{<https://lammps.sandia.gov/>}.

\bibitem{woodParker1957}
W. W. Wood and F. R. Parker, ``Monte Carlo equation of state of molecules interacting with the Lennard-Jones potential. I. a supercritical isotherm at about twice the critical temperature,'' J. Chem. Phys. \textbf{27}, 720--733 (1957).

\bibitem{kobAndersenPRL1994}
W. Kob and H. C. Andersen, ``Scaling behavior in the $\beta$-relaxation regime of a supercooled Lennard-Jones mixture,'' Phys. Rev. Lett. \textbf{73}, 1376--1379 (1994).

\bibitem{guzmanDePablo2003} O. Guzm\'an and J. J. de Pablo, ``An effective-colloid pair potential for Lennard-Jones colloid-polymer mixtures,'' J. Chem. Phys. \textbf{118}, 2392--2397 (2003).

\bibitem{bennemann1998}Ch. Bennemann, W. Paul, and K. Binder, ``Molecular-dynamics
simulations of the thermal glass transition in polymer melts: $\alpha$-relaxation behavior,'' Phys. Rev. E \textbf{57}, 843--851 (1998).

\bibitem{HansenVerlet1969}J.-P. Hansen and L. Verlet, ``Phase transitions of the Lennard-Jones system, '' Phys. Rev. \textbf{184}, 151--161 (1969).

\bibitem{abramo2015}M. C. Abramo, C. Caccamo, D. Costa, P. V. Giaquinta, G. Malescio, G. Muna\`o, and S. Prestipino, ``On the determination of phase boundaries via thermodynamic integration across coexistence regions,'' J. Chem. Phys. \textbf{142}, 214502-1--10 (2015).

\bibitem{HansenMcDonald}J.-P. Hansen and I. R. McDonald, \emph{Theory of Simple
Liquids: With Applications to Soft Matter} (Academic Press, Boston, 2013).

\bibitem{allen90}M. P. Allen and D. J. Tildesley, \emph{Computer Simulation of Liquids} (Oxford University Press, New York, 1990).

\bibitem{rapaport98}D. C. Rapaport, \emph{The Art of Molecular Dynamics Simulation} (Cambridge University Press, Cambridge, UK, 2002).

\bibitem{kobAndersenPRE51_1995}W. Kob and H. C. Andersen, ``Testing made-coupling theory for a supercooled binary Lennard-Jones I: The van Hove correlation function,'' 
Phys. Rev. E \textbf{51}, 4626--4641 (1995).

\bibitem{kobAndersenPRE52_1995}W. Kob and H. C. Andersen, ``Testing
mode-coupling theory for a supercooled binary Lennard-Jones mixture. II.
Intermediate scattering function and dynamic susceptibility,'' Phys. Rev. E \textbf{52}, 4134--4153 (1995).

\bibitem{KVLKobBinder_1996}K. Vollmayr, W. Kob, and K. Binder, ``How do the properties of a glass depend on the cooling rate? A computer simulation study of a Lennard-Jones system,'' J. Chem. Phys. \textbf{105}, 4714--4728 (1996).

\bibitem{hassani2016}M. Hassani, P. Engels, D. Raabe, and F. Varnik, ``Localized plastic deformation in a model metallic glass: a survey of free volume and local force distributions,'' J. Stat. Mech.: Theory Exp. . 084006-1--11 (2016).

\bibitem{schoenholz2016}
S. S. Schoenholz, E. D. Cubuk, D. M. Sussman, E. Kaziras, and A. J. Liu,
``A structural approach to relaxation in glassy liquids,'' Nat. Phys. \textbf{12}, 469--471 (2016).

\bibitem{ShrivastavPRE2016}
G. P. Shrivastav, P. Chaudhuri, and J. Horbach, ``Yielding of
glass under shear: A direct percolation transition precedes shear-band formation,''
Phys. Rev. E \textbf{94}, 042605-1--10 (2016).

\bibitem{makeevPriezjev2018}M. A. Makeev and N. V. Priezjev, ``Distributions of pore sizes and atomic densities in binary mixtures revealed by molecular dynamics simulations,''
Phys. Rev. E \textbf{97}, 023002-1--8 (2018).

\bibitem{pedersenSchroederDyrePRL_2018}
U. R. Pedersen, Th. B. Schr\o der, and J. C. Dyre, ``Phase diagram of Kob-Andersen-type binary Lennard-Jones mixtures,'' Phys. Rev. Lett. \textbf{120}, 165501-1--5 (2018).

\bibitem{gouldTobochnikChristian2007}
H. Gould, J. Tobochnik, and W. Christian, \emph{An Introduction to Computer Simulation Methods: Applications to Physical Systems} (Pearson: Addison Wesley, San Francisco, 2007).

\bibitem{newman2013}
M. Newman, \emph{Computational Physics} (Createspace, North Charleston, 2013).

\bibitem{numrecipes}
W. H. Press, S. A. Teukolsky, W. T. Vetterling, and B. P. Flannery, \emph{Numerical Recipes: The Art of Scientific Computing}, 3rd ed. (Cambridge University Press, New York, 2007).

\bibitem{schroeder2000}
D. V. Schroeder, \emph{An Introduction to Thermal Physics} (Addison-Wesley Longman, San Francisco, 2000).

\bibitem{blundells}S. J. Blundell and K. M. Blundell, \emph{Concepts in Thermal Physics} (Oxford University Press, New York, 2010).

\bibitem{gouldTobochnikStatMech}
H. Gould and J. Tobochnik, \emph{Statistical and Thermal Physics} (Princeton University Press, Princeton, 2010).

\bibitem{martynaMolPhys1996}
G. J. Martyna, M. E. Tuckerman, D. J. Tobias, and M. L. Klein, ``Explicit reversible integrators for extended systems dynamics,'' Mol. Phys. \textbf{87}, 1117--1157 (1996).

\bibitem{andersen1980}H. C. Andersen, ``Molecular dynamics simulations at constantpressure and/or temperature,'' J. Chem. Phys. \textbf{72}, 2384--2393 (1980).

\bibitem{andrea}T. A. Andrea, W. C. Swope, and H. C. Andersen, ``The role of long ranged forces in determining the structure and properties of liquid water,'' J. Chem. Phys. \textbf{79}, 4576--4584 (1983).

\bibitem{LAMMPSfixnvtp}LAMMPS documentation for the fix NVT command, \url{<https://lammps.sandia.gov/doc/fix_nh.html>}.

\bibitem{nosehoover}W. G. Hoover, ``Canonical dynamics: Equilibrium phase-space distributions,'' Phys. Rev. A \textbf{31}, 1695--1697 (1985).

\bibitem{Taylor}J. R. Taylor, \emph{Classical Mechanics} (University Science Books, Sausalito, 2005).

\bibitem{Frenkel2002}
D. Frenkel and B. Smit, \emph{Understanding Molecular Simulation: From Algorithms to Applications} (Academic Press, San Diego, 2002).

\bibitem{nose1984}
S. Nos\'e, ``A molecular dynamics method for simulations in the canonical ensemble,'' Mol. Phys. \textbf{52}, 255--268 (1984).

\bibitem{branka}
A. C. Bra\'nka and K. W. Wojciechowski, ``Generalization of Nos\'e and Nos\'e-Hoover isothermal dynamics,'' Phys. Rev. E \textbf{62}, 3281--3292 (2000).

\bibitem{martynaJCP1994}
G. J. Martyna, D. J. Tobias, and M. L. Klein, ``Constant pressure molecular dynamics algorithms,'' J. Chem. Phys. \textbf{101}, 4177--4189 (1994).

\bibitem{tuckerman2001}
M. E. Tuckerman, Y. Liu, G. Ciccotti, and G. J. Martyna, ``Non-Hamiltonian molecular dynamics: Generalizing Hamiltonian phase space principles to non-Hamiltonian systems,'' J. Chem. Phys. \textbf{115}, 1678--1702 (2001).

\bibitem{shinoda2004}
W. Sinoda, M. Shiga, and M. Mikami, ``Rapid estimation of elastic constants by molecular dynamics simulation under constant stress,'' Phys. Rev. B \textbf{69}, 134103-1--8 (2004).

\bibitem{martynaKleinTuckermanJCP1992}
G. J. Martyna, M. L. Klein, and M. Tuckerman, ``Nos\'e-Hoover chains: The
canonical ensemble via continuous dynamics,'' J. Chem. Phys. \textbf{97}, 2635--2643 (1992).

\bibitem{fox}
J. R. Fox and H. C. Andersen, ``Molecular dynamics simulations of a supercooled monatomic liquid and glass,'' J. Phys. Chem. \textbf{88}, 4019--4027 (1984).

\bibitem{tapiasJPC2016}
D. Tapias, D. P. Sanders, and A. Bravetti, 
``Geometric integrator for simulations in the canonical ensemble,'' J. Chem. Phys. \textbf{145}, 084113-1--9 (2016).

\bibitem{newmanOnline}
M. Newman, Computational physics, \url{<http://www-personal.umich.edu/ mejn/cp/>}, accessed: 2019-09-04.

\bibitem{supplement}
Python and LAMMPS scripts are in the ancillary files of the arXiv
version of this manusscript.

\bibitem{LAMMPSdownload}Download the source and documentation as a tarball at  \url{<https://lammps.sandia.gov/doc/Install_tarball.html>}.

\bibitem{LAMMPSvariable}
See the LAMMPS documentation at \url{<https://lammps.sandia.gov/doc/variable.html>}.

\bibitem{kobBarratPRL1997}
W. Kob and J.-L. Barrat, ``Aging effects in a Lennard-Jones glass,'' Phys. Rev. Lett. \textbf{78}, 4581--4584 (1997).

\bibitem{kobBarratPhysicaA1999}
W. Kob and J.-L. Barrat, ``Aging in a Lennard-Jones glass,'' Physica A \textbf{263}, 
234--241 (1999).

\bibitem{kobetalJPhysCondMat2000}
W. Kob, J.-L. Barrat, F. Sciortino, and P. Tartaglia,
``Aging in a simple glass former,'' J. Phys.: Condens. Matter \textbf{12}, 6385--6394
(2000).

\bibitem{LAMMPSfixavetime}See the LAMMPS  documentation at \url{<https://lammps.sandia.gov/doc/fix_ave_time.html>}.

\bibitem{logfreq3}{\tt logfreq3} was added to LAMMPS in June 2019. The example {\tt logfreq3(10,25,1000)} is explained at Ref.~\onlinecite{LAMMPSvariable}.

\bibitem{logpython}To reproduce these   times with logarithmic time using Python, choose $\tt t0msd=10$, $\tt nmsdmax=24$, and in the time loop choose {\tt tmsdnextint = ceil(tmsd)} instead of
{\tt tmsdnextint=int(round(tmsd))}.

\bibitem{durianPRL1995}
D. J. Durian, ``Foam mechanics at the bubble scale,''
Phys. Rev. Lett. \textbf{75}, 4780--4783 (1995).

\bibitem{durianPRE1997}
D. J. Durian, ``Bubble-scale model of foam mechanics: Melting, nonlinear behavior, and avalanches,'' Phys. Rev. E \textbf{55}, 1739--1751 (1997).

\bibitem{DKAFig3}W. Kob, private communication.

\bibitem{LAMMPS2d}See the LAMMPS documentation for 
2d simulations at \url{<https://lammps.sandia.gov/doc/Howto_2d.html>}.

\bibitem{KALJ2dBruening}
R. Br\"uning, D. A. St-Onge, S. Patterson, and W. Kob,
``Glass transitions in one-, two-, three-, and four-dimensional binary Lennard-Jones systems,'' J. Phys.: Condens. Matter \textbf{21}, 035117-1--11 (2009).

\bibitem{KVLRomanHorbach}
K. Vollmayr-Lee, J. A. Roman, and J. Horbach, ``Aging to equilibrium dynamics of SiO$_2$,'' 
Phys. Rev. E \textbf{81}, 061203-1-9 (2010).

\end{thebibliography}
\end{document}